\documentclass[journal,twocolumn,twoside]{IEEEtran}
\pdfoutput=1
\usepackage{array,algpseudocode,algorithmicx,algorithm,amssymb,bm,paralist,balance,multirow,import,url}
\usepackage[caption=false,font=footnotesize]{subfig}
\usepackage{cite}
\usepackage[cmex10]{amsmath}
\usepackage[pdftex]{graphicx}
\usepackage[space]{grffile}
\usepackage{xcolor}
\usepackage{cleveref}
\usepackage[inline]{enumitem}
\setlist[enumerate,1]{label={(\alph*)}}

\crefname{myThm}{Theorem}{Theorems}

\crefname{myDef}{Definition}{Definitions}
\newtheorem{myLem}{Lemma}
\crefname{myLem}{Lemma}{Lemmas}
\newtheorem{myProp}{Proposition}
\crefname{myProp}{Proposition}{Propositions}
\newtheorem{rem}{Remark}
\crefname{rem}{Remark}{Remarks}

\crefname{myCor}{Corollary}{Corollaries}
\newtheorem{myProb}{Problem}
\crefname{myProb}{Problem}{Problems}

\crefname{myExample}{Example}{Examples}

\newcommand{\defeq}{\stackrel{\text{\tiny{def}}}{=}}

\newcommand{\imagj}{\mathrm{j}}

\def\Vecf{\mathbf{f}}

\def\Vecp{\mathbf{p}}
\def\Vecq{\mathbf{q}}

\def\Vecs{\mathbf{s}}

\def\Vecw{\mathbf{w}}
\def\Vecx{\mathbf{x}}
\def\Vecy{\mathbf{y}}

\def\Veckappa{\bm{\kappa}}

\def\VecD{\mathbf{D}}

\def\VecF{\mathbf{F}}
\def\VecG{\mathbf{G}}
\def\VecI{\mathbf{I}}

\def\VecK{\mathbf{K}}
\def\VecH{\mathbf{H}}

\def\VecP{\mathbf{P}}

\def\VecT{\mathbf{T}}
\def\VecU{\mathbf{U}}

\def\VecIndexn{\bm{n}}
\def\VecIndexk{\bm{k}}
\def\VecIndexj{\bm{j}}
\def\VecIndexL{\bm{L}}

\def\Elltwo{\mathcal{L}^2}

\DeclareMathOperator{\SPAN}{span}

\def\transpose{^{\!\mathsf{T}}}

\allowdisplaybreaks

\DeclareTextCommandDefault{\textcopyright}{\textcircled{c}}

\newcommand{\MyBIBLocation}{./Bibliography}

\begin{document}
\pdfsuppresswarningpagegroup=1
\title{A Sampling Framework for Solving Physics-driven Inverse Source Problems}

\author{John~Murray-Bruce*,~\IEEEmembership{Member,~IEEE,}
        and~Pier~Luigi~Dragotti,~\IEEEmembership{Fellow,~IEEE}
\thanks{J. Murray-Bruce is with the Department of Electrical and Computer Engineering, Boston University, MA, 02215, USA, (e-mail: johnmb@bu.edu) but conducted the work in this manuscript whilst with the Communications and Signal Processing Group at Imperial College, London.}
\thanks{P. L. Dragotti is with the Communications and Signal Processing Group, Department of Electrical and Electronic Engineering, Imperial College, London, United Kingdom, SW7 2AZ (e-mail:p.dragotti@imperial.ac.uk).}
\thanks{This work is supported by the European Research Council (ERC) starting investigator award Nr. 277800 (RecoSamp).}%
\thanks{Some of the work in this paper has, in part, been presented in \cite{Murray1608_Solving,Murray1703_Solving}.}%
}

\markboth{IEEE Transactions on Signal Processing}{Murray-Bruce and Dragotti: Solving Physics-driven Inverse Source Problems}

\maketitle

\begin{abstract}
Partial differential equations are central to describing many physical phenomena. In many applications these phenomena are observed through a sensor network, with the aim of inferring their underlying properties. Leveraging from certain results in sampling and approximation theory, we present a new framework for solving a class of inverse source problems for physical fields governed by linear partial differential equations. Specifically, we demonstrate that the unknown field sources can be recovered from a sequence of, so called, generalised measurements by using multidimensional frequency estimation techniques. Next we show that---for physics-driven fields---this sequence of generalised measurements can be estimated by computing a linear weighted-sum of the sensor measurements; whereby the exact weights (of the sums) correspond to those that reproduce multidimensional exponentials, when used to linearly combine translates of a particular prototype function related to the Green's function of our underlying field. Explicit formulae are then derived for the sequence of weights, that map sensor samples to the exact sequence of generalised measurements when the Green's function satisfies the generalised Strang-Fix condition. Otherwise, the same mapping yields a close approximation of the generalised measurements. Based on this new framework we develop practical, noise robust, sensor network strategies for solving the inverse source problem, and then present numerical simulation results to verify their performance.
\end{abstract}

\begin{IEEEkeywords}
Partial differential equations (PDEs), inverse problems, universal sampling, sensor networks, diffusion equation, wave equation, Strang-Fix conditions, Prony's method.
\end{IEEEkeywords}

%
\IEEEpeerreviewmaketitle


\section{Introduction}
\label{sec:Introduction}
\IEEEPARstart{S}{ensor networks}, and the use thereof, for sensing and monitoring physical fields is receiving significant research attention due, in part, to the significant advances made over the last few decades in the fields of (wireless) networking, communications and in the fabrication of microprocessors \cite{Akyildiz02_survey,Kumar2002_collaborative}. During this period many interesting applications in localisation, tracking and parameter estimation have been considered \cite{Marano2008_distributed,Braca2008_Enforcing}. The sensor nodes are deployed over a region of interest to obtain spatiotemporal samples of some physical phenomena. Often, these phenomena are driven by natural mechanisms that typically involve the transportation of matter/particles or the transportation of energy, from one point to another and thus can be described by partial differential equations (PDEs). For example, the mode of transport governing the dispersion of plumes in environmental monitoring \cite{Ranieri2012_Sampling}, spreading of fungal diseases in precision agriculture \cite{Langendoen2006_Potatoes}, biochemical and nuclear wastes \cite{Matthes2005_Source} is well-known to be diffusion and the corresponding diffusion field is the variation in concentration of the released substance over space and time.

Besides diffusion, there exist numerous modes of transport and corresponding fields, such as wave and potential fields that have also received considerable research attention from the signal processing community. Such efforts have focussed on developing robust sensor data fusion schemes that either infer the sources inducing the measured field or directly reconstruct the field. Often these physical fields of interest are spatially non-bandlimited and so require an extremely dense set of samples in order to achieve a faithful recovery, using the classical linear bandlimited (BL) reconstruction framework.

In the localisation of neuronal source activities from electroencephalographic (EEG) signals \cite{Becker2015_Brain,Grech2008_Review}---for bioengineering applications---the use of Poisson's equation to model the brain activity is important, since this PDE accurately describes the relationship between the measured electrical potentials (the field) and the current dipoles (the sources). To alleviate some of the limitations of a BL reconstruction when solving this EEG-related inverse source problem (ISP), many approaches utilising least-squares \cite{Dale2000_Dynamic}, sparsity-based \cite{Matsuura1997_Robust,Lin2006_Distributed} and, more recently, cosparsity-based \cite{Albera2014_Brain,Kitic2016_Physics} regularisation have been proposed. Moreover, Bayesian modelling \cite{Wipf2010_Robust}, beamforming (see \cite{Limpiti2006_Cortical,Piotrowski2014_Reduced} and references therein), as well as subspace techniques \cite{Chevalier2006_HighResolution} have also been explored.

Wave fields are also prevalent in applications such as acoustic tomography \cite{Jovanovic2009_Acoustic}, speech and sound enhancement \cite{Chen2002_Source}, sound/wave source localisation \cite{Chen2002_Source} and many more. In such situations the wave equation provides a physical law that describes the propagation of such fields. To solve the associated ISPs, several interesting techniques have been put forward; the classical and most commonly used techniques are based on maximum likelihood estimation and beamforming \cite{Chen2003_Acoustic}. Recently however, Dokmani\`{c} \textit{et al} \cite{Dokmani2016_Acoustic,Dokmanic2015_Listening} proposed an approach that exploits certain salient properties of euclidean distance matrices to solve the simultaneous localisation and mapping problem for acoustic sources. Kiti\'{c} \textit{et al} \cite{Kitic2016_Physics} also formulated an optimisation problem by using the cosparse regularisation framework, whilst a finite rate of innovation (FRI) based method is introduced in \cite{Dogan2014_Finite} to solve this acoustic ISP using boundary-only measurements.

For the diffusion field reconstruction problem, Reise \textit{et al} \cite{Reise2012_Distributed} propose the use of \textit{hybrid shift-invariant subspaces}, whilst recovery algorithms based on the use of finite element methods (FEMs) \cite{vanWaterschoot2011_Static,vanWaterschoot2012_Distributed} and compressive sensing (CS) \cite{Ranieri2011_Samp,Paul2016_Sparse} have also been proposed. Moreover statistical estimation techniques, based on Bayesian estimation and Kalman filtering, have also been studied, see \cite{Nehorai95_Detection,Weimer2009_Multiple,Sawo2006_Bayesian} for instance. Meanwhile, Dokmanic \textit{et al} \cite{Dokmanic2011_Sensor} retrieve the single source parameters by approximating the resulting field using a truncated Fourier series, and Lu \textit{et al} demonstrate that by solving a set of linear equations the single diffusion source parameters can be estimated \cite{Lu2011a_Localizing}. 
Furthermore, in \cite{Murray1506_Estimating} we showed that given a proper sequence of generalised measurements it is possible to recover the unknown parameters for a specific class of diffusion source distributions. We will leverage from that concept, in this current work, to devise a framework for solving a more \textit{general} class of physics-driven ISPs. Specifically, the new approach will apply to higher dimensional ISPs governed by a larger class of PDE models.%

The rest of this paper is organised as follows. In \Cref{sec:IPFormulation}, we give mathematical descriptions of some common linear, constant coefficient PDEs encountered in many applications and then state the related ISP and also discuss the sensor network model. In \Cref{sec:SensingFunctions} we outline how to solve the ISP of interest assuming we have access to a set of generalised measurements. Specifically, we discuss and state explicitly how to select properly the spatiotemporal sensing functions in order to be able to solve the $d$-dimensional ISP. Then we explore a new approach for computing the desired generalised measurements in \Cref{sec:SensorData2GeneralizedMeasurements}, based on taking proper linear combinations of the sensor data. In particular we realise that this leads to the well-known exponential reproduction problem---using translates of a prototype function---encountered in approximation theory and in the FRI framework \cite{Uriguen2013FRI}. Here however, the prototype function coincides with the space- and time-reversed Green's function of the physical field. Then we derive conditions on the Green's function, for which this exponential reproduction problem can be exactly or approximately solved. \Cref{sec:SNAlgorithms}, discusses how to adapt this new framework to solving ISPs using sensor networks. We develop explicit centralised and distributed estimation strategies whilst considering both uniform and nonuniform sensor placements. Then in \Cref{sec:NumericalResults}, we provide numerical simulation results to validate the new framework, comparing it against a sparsity-based recovery method; and finally, we conclude the paper in \Cref{sec:Conclusion}.

\section{Physics-driven inverse problems: Problem formulation}
\label{sec:IPFormulation}
The term \textit{physics-driven} is used in this paper to describe physical phenomena, specifically physical fields, that propagate through space and time according to some linear partial differential equation (PDE). In a more general form, such phenomena can be written as,
\begin{equation}
\mathcal{D} u(\Vecx,t) = f(\Vecx,t),
\label{eq:GeneralForm_LinearPDE}
\end{equation}
where $\mathcal{D}$ denotes a linear differential operator, whilst $f(\Vecx,t)$ is the source of the field $u(\Vecx,t)$ propagating through space $\Vecx \in \Omega \subset \mathbb{R}^d$ and time $t \in \mathbb{R}_{+}$. According to the method of Green's functions, under certain boundary conditions, the system \eqref{eq:GeneralForm_LinearPDE} admits the solution
\begin{equation}
	u(\Vecx,t) = (g \ast f)(\Vecx,t),
\label{eq:Convolution_GreensFunc}
\end{equation}
where $g(\Vecx,t)$ is the so called Green's function of the underlying field. The following fields will be considered in this paper:
\begin{enumerate}
		\item \textbf{Potential fields:} these are encountered frequently in many situations arising in electrostatics. Mathematically, the potential field satisfies
		\begin{equation}
			\nabla^2 u(\Vecx) = f(\Vecx).
		\label{eq:PoissonEqn}
		\end{equation}		
		The Green's function for this PDE in 2D (i.e. $d=2$) is,
		\begin{equation}
			g(\Vecx) = \frac{1}{2 \pi} \log \mathopen{}\left( \| \Vecx \| \right),
		\label{eq:GreensFunction_Poisson2D}
		\end{equation}
		whilst for $d=3$ the Green's function becomes
		\begin{equation}
			g(\Vecx) = -\frac{1}{4 \pi \| \Vecx \|}.
		\label{eq:GreensFunction_Poisson3D}
		\end{equation}
		\item \textbf{Diffusion fields:} refer to physical phenomena such as the propagation of heat, plumes and leakages that can be described mathematically by,
		\begin{equation}
			\frac{\partial}{\partial t}u(\Vecx, \! t) = \mu \nabla^2 u(\Vecx,\! t) {+} f(\Vecx,\! t),
		\label{eq:DiffusionEqn}
		\end{equation}
		with the following Green's function, for $d \geq 1$:
		\begin{equation}
			g(\Vecx, t) = \frac{1}{(4 \pi \mu t)^{d/2}} e^{- \frac{\left\|\Vecx\right\|^2}{4 \mu t}} H(t),
		\label{eq:GreensFunction_Diff}
		\end{equation}
		where $H(t)$ is the unit step and $\mu$ is the diffusivity.
		\item \textbf{Wave fields:} describe many situations arising for example in acoustics and electromagnetism. The wave equation is given by:
		\begin{equation}
			\nabla^2 u(\Vecx, t) - \frac1{c^2} \frac{\partial^2}{\partial t^2}u(\Vecx, t) = f(\Vecx, t),
		\label{eq:WaveEqn}
		\end{equation}
		where the wave field $u(\Vecx,t)$ induced by the source distribution $f(\Vecx, t)$ propagates through the medium at a speed $c$. The Green's function for the $2$-D wave equation (i.e. $d = 2$) is given by:
		\begin{equation}
			g(\Vecx,t) = \frac{c}{2 \pi \sqrt{c^2 t^2 - \| \Vecx \|^2}} H(ct - \| \Vecx \|).
		\label{eq:GreensFunction_Wave2D}
		\end{equation}
		Moreover, in $3$-D (i.e. $d = 3$), it can be shown that:
		\begin{equation}
			g(\Vecx,t) = \frac{1}{4 \pi \| \Vecx \|} \delta(t - \| \Vecx \|/c).
		\label{eq:GreensFunction_Wave3D}
		\end{equation}
\end{enumerate}

The Green's functions above assume a Sommerfeld radiation condition -- i.e. a \textit{quiescent condition} at an initial time, such that $\left. u(\Vecx, t)\right|_{t=0} = \left. \frac{\partial}{\partial t}u(\Vecx, t)\right|_{t=0} = 0$ and a \textit{convergence condition} at infinity, meaning $\left. u(\Vecx, t)\right|_{\|\Vecx\| \rightarrow \infty} = \left. \frac{\partial}{\partial x_1}u(\Vecx, t)\right|_{\|\Vecx\| \rightarrow \infty} = \left. \frac{\partial}{\partial x_2}u(\Vecx, t)\right|_{\|\Vecx\| \rightarrow \infty} = 0$. See for example, \cite{Duffy2001_Green} for a derivation of these expressions.

Given the PDE models of such fields, the aim of this paper is to develop a framework for solving the associated ISP, from spatiotemporal sensor network measurements of the induced field\footnote{For instance using a suitable microphone array for audio fields, or an array of thermal sensors to monitor the temperature of a room.}. Precisely, the ISP considered here is the following:
\begin{myProb}
Let $\mathcal{S} = \{\Vecx_n\}_{n=1}^N$ denote a network of $N$ sensors, so that the $n$-th sensor situated at $\Vecx_n$ collects samples $\varphi_{n}(t_l) =u(\Vecx_n, t_l)$ of the field $u$, at times $t_l$ for $l = 0,1,\ldots,L$. Given these spatiotemporal samples and knowledge of the Green's function of the field, we intend to estimate the unknown source distribution $f(\Vecx,t)$.
\end{myProb}%
\subsection{Sensor Network Model and Assumptions}
\label{ssec:SensorNetworkAssumptions}
The sensor networks used to monitor our physics-driven fields are such that:
\begin{enumerate}
	\item They comprise $N$ sensor nodes deployed (uniformly or randomly) over the region of interest ($\Omega \subset \mathbb{R}^d$). For example in $2$-D the sensors all lie in the same plane.
	\item The sensor locations $\Vecx_n \in \Omega$ are known and each sensor samples the field locally at time instants $t_l$ for $l=0,1,\ldots,L$. Hence the noiseless field samples are simply the field \eqref{eq:Convolution_GreensFunc}, evaluated at $\Vecx = \Vecx_n$ and $t = t_l$ as follows:
\begin{equation}
\varphi_n(t_l) = u(\Vecx_n,t_l).
\label{eq:NoiselessSamples}
\end{equation}
	\item The sensor noise can be modelled by a zero mean additive white Gaussian noise (AWGN) process, so that the noisy measurements are:
	\begin{equation}
		\varphi_{n,l}^{\epsilon} = \varphi_{n}(t_l) + \epsilon_{n,l},
	\label{eq:NoisySamples}
	\end{equation}
	where $\epsilon_{n,l} \sim \mathcal{N}(0,\sigma^2)$ and the (average) signal-to-noise ratio ($\mathrm{SNR}$) is
\begin{equation}
	\mathrm{SNR} \defeq 10 \log_{10}\left( \frac{\sum_{n=1}^N \sum_{l=0}^{L} |\varphi_{n}(t_l)|^2 }{ N(L+1)\sigma^2}\right).
	\label{eq:SNR_model}
\end{equation}
	\item The sensor nodes are synchronised. Hence the sensors sample the field at the same instants.%
\end{enumerate}

Within this setting, we consider two scenarios: a \textit{centralised} scenario where we assume that all sensors' readings are available at a fusion centre at which all the processing is performed and a \textit{distributed} scenario where sensors can perform processing locally but can only communicate with neighbouring sensors. In this distributed setup, the sensors must recover the unknown distribution $f(\Vecx,t)$ through localised/in-network data processing and communications alone.

\begin{figure}[ht]
  \centering 
  \includegraphics[trim = 0mm 11.5mm 0mm 8mm, clip, width=0.7\linewidth]{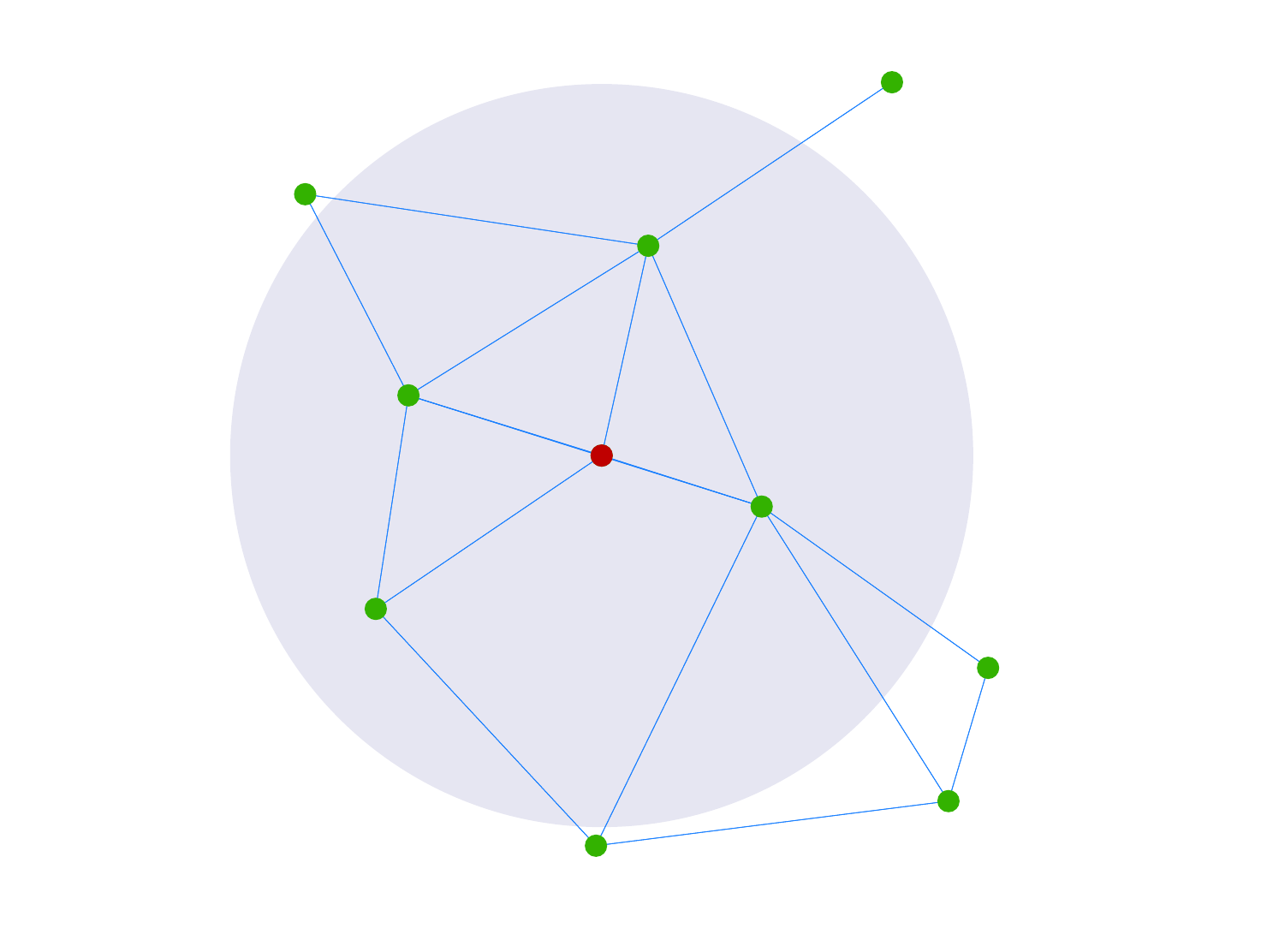}
	\caption{\textbf{Sensor network.} A distributed sensor network as modelled by a RGG.}
\label{fig:RGG_links}
\end{figure}

To model the network topology, in the distributed setting, we assume a \textit{connected} random geometric graph (RGG), denoted by $\mathcal{G}(N,r_{\mathrm{con}})$, with $N$ sensor nodes and connectivity radius $r_{\mathrm{con}}$. To realise this, we place $N$ nodes uniformly at random over a unit square/cube and then put an edge between a pair of nodes if their Euclidean distance is at most $r_{\mathrm{con}}$. An example, with $N=10$ nodes is shown in \Cref{fig:RGG_links}, the shaded circular region is the communication radius of the red sensor.
%
%
%
%

\section{Choosing the sensing functions: Source recovery from generalised measurements}
\label{sec:SensingFunctions}
The focus of this paper will be predominantly on fields induced by multiple localised and instantaneous sources. This source distribution $f(\Vecx,t)$ admits the parametrisation:
\begin{equation}%
	f(\Vecx, t) = \displaystyle \sum_{m=1}^{M} c_m {\delta(\Vecx-\bm{\xi}_m, t - \tau_m)},
\label{eq:InstSourceDist}
\end{equation}
where $M$ is the total number of sources, $c_m, \tau_m \in \mathbb{R}$ and $\bm{\xi}_m = (\xi_{i,m})_{i=1}^d \in \mathbb{R}^d$ are the intensity, activation time and location of the $m$-th source respectively. %
As a result of this parametrisation, the ISP becomes one of estimating $M$ triples $\{ (c_m, \tau_m, \bm{\xi}_m) \}_{m=1}^M$. 

The proposed framework will be based on estimating the unknown source parameters from the following multidimensional sequence of generalised measurements,
\begin{equation}
\mathcal{Q}(\VecIndexk,r) = \left\langle f(\Vecx,t), \Psi_{\VecIndexk}(\Vecx)\Gamma_r(t)  \right\rangle_{\Vecx,t},
\label{eq:GeneralizedMeasurementsVec}
\end{equation}
where $\{\Psi_{\VecIndexk}(\Vecx)\}_{\VecIndexk \in \mathbb{N}^d}$ and $\{\Gamma_r(t)\}_{r \in \mathbb{N}}$, are families of properly chosen \textit{spatial} and \textit{temporal sensing functions} respectively. We will discuss how to choose these in the sequel.

Moreover, observe that when we substitute \eqref{eq:InstSourceDist} into the inner product \eqref{eq:GeneralizedMeasurementsVec}, we obtain:
\begin{equation}
	\mathcal{Q}(\VecIndexk,r) = \sum_{m=1}^M c_m \Psi_{\VecIndexk}(\bm{\xi}_m) \Gamma_r(\tau_m).
\label{eq:GeneralizedMeasurements_STSF_Relation}
\end{equation}
Hence, our first task will be to select $\Psi_{\VecIndexk}(\Vecx)$ and $\Gamma_r(t)$ so that we can recover $\{ c_m, \tau_m, \bm{\xi}_m \}_{m=1}^M$ from $\{\mathcal{Q}(\VecIndexk,r)\}_{\VecIndexk,r}$. We then discuss in \Cref{sec:SNAlgorithms} how to obtain the measurements \eqref{eq:GeneralizedMeasurementsVec} from the sensors' readings. In the same spirit of \cite{Murray1506_Estimating}, we propose the use of exponentials with purely imaginary exponents as our sensing functions for two main reasons. The first is that, the sum \eqref{eq:GeneralizedMeasurements_STSF_Relation} becomes a superposition of multidimensional exponentials, i.e. a multidimensional system of superimposed sinusoids, which can be efficiently solved by using multidimensional extensions of Prony's frequency estimation methods \cite{Hua1990_Matrix,Maravic2004_Exact,Sahnoun2016_Multidimensional}. The second reason is because the use of imaginary exponentials improves the stability of the estimation problem, since in this case the magnitude of the terms in the generalised sequence remain bounded. We now examine explicitly the $d$-dimensional source recovery problem. Note however that, in most natural applications, one would mostly be interested in two- and three-dimensional fields.
\subsection{Sensing sources in space and time}
The temporal sensing function is chosen to be the exponential $\Gamma_r(t) = e^{\imagj r t/T}$, where $T = t_L$ is the instant at which the sensors measure the last sample of the field. In addition, we choose $\Psi_{\VecIndexk}(\Vecx) = e^{  \imagj \VecIndexk \cdot \Vecx}$ with $\VecIndexk \defeq (k_1,k_2, \ldots, k_d) \in \mathbb{N}^d$. Under this particular selection of spatial and temporal sensing functions, the expression \eqref{eq:GeneralizedMeasurements_STSF_Relation} becomes:
\begin{equation}
	\mathcal{Q}(\VecIndexk,r) = \sum_{m=1}^M c_m e^{\imagj r \tau_m/T} e^{  \imagj k_1 \xi_{1,m} + \imagj k_2 \xi_{2,m}+ \cdots + \imagj k_d \xi_{d,m}}.
\label{eq:PowerSumSeries_ND}
\end{equation}

Notice now that, for some fixed $r \neq 0$ (take, for instance, the particular case $r = 1$) then expression \eqref{eq:PowerSumSeries_ND} is of the form:
\begin{equation*}
 \mathcal{Q}(\VecIndexk,1) \defeq \mathcal{Q}(k_1,k_2,\ldots, k_d, 1) = \sum_{m=1}^M b_m \prod_{i=1}^d v_{i,m}^{k_i},
\end{equation*}
with $b_m {=} c_m e^{\imagj \tau_m/T}$ and $v_{i,m} {=} e^{  \imagj \xi_{i,m}}$. This is a multidimensional Prony-like system which, as described in Appendix \ref{app:ParamEstimation_SoS}, can be solved to obtain $\{ (b_m, v_{1,m}, v_{2,m}, \ldots, v_{d,m})\}_{m=1}^M$ simultaneously from $\{\mathcal{Q}(\VecIndexk,1)\}_{\VecIndexk}$ provided $k_i {=} 0,1,\ldots,K_i$ and $K_i {\geq} 2M {-} 1$ for any $i \in {1,2,\ldots,d}$. See also \cite{Maravic2004_Exact,Hua1990_Matrix}, for more on the topic of multidimensional frequency estimation. Then it follows immediately that the unknown source parameters are given by: $c_m = |b_m|$, $\tau_m = T \arg(b_m)$ and $\bm{\xi}_m = -\imagj\left( \log(v_{1,m}), \log(v_{2,m}), \ldots, \log(v_{d,m}) \right)$.
Having outlined how to recover the unknown point source parameters from $\{\mathcal{Q}(\VecIndexk,r)\}_{\VecIndexk}$, we must now focus on retrieving these generalised measurements from the spatiotemporal samples of the field.

In \cite{Murray1506_Estimating,Murray1602_Physics} it was shown that for the case of the diffusion field, the generalised measurements can be found by imposing that $\Psi_k(\Vecx)$ be analytic. A similar strategy has been used in \cite{Dogan2014_Finite,DoganFRI2012} for the wave and Poisson equation. The disadvantage of these approaches is that \begin{inparaenum}[\itshape(i)] \item they cannot be easily extended to $d > 2$ and \item the constraint that $\Psi_k$ is analytic leads in some cases to less stable reconstruction algorithms. \end{inparaenum}%
Consequently, in the following section we outline a new and more versatile approach to compute $\mathcal{Q}(\VecIndexk,r)$ from the sensor data.%

Notice that if we allow $k_1$ to be imaginary and impose, $k_1 = -\imagj k_2 = \imagj k$ we obtain the same $\mathcal{Q}(k,r)$ of \cite{Murray1506_Estimating}, so this new approach is a generalisation.

\section{Multidimensional generalised measurements from sensor samples}
\label{sec:SensorData2GeneralizedMeasurements}
To compute the desired set of multidimensional generalised measurements we consider taking weighted linear combinations of the spatiotemporal samples:
\begin{equation}
	\sum_{n=1}^N \sum_{l=0}^L w_{n,l} (\VecIndexk,r) \varphi_{n}(t_l) = \widehat{\mathcal{Q}}(\VecIndexk,r),
	\label{eq:WeightedSummationGeneralizedMeasurements}
\end{equation}
hence the goal is to find the weights $\{w_{n,l} (\VecIndexk,r)\}_{n,l}$, such that the left hand side of \eqref{eq:WeightedSummationGeneralizedMeasurements}, that is $\widehat{\mathcal{Q}}(\VecIndexk,r)$, coincides with the right hand side of \eqref{eq:GeneralizedMeasurementsVec} for any $\VecIndexk$ and $r$. The result below follows from this consideration.

\begin{myProp}
Computing the multidimensional sequence of generalised measurements $\{ \mathcal{Q}(\VecIndexk,r) \}_{\VecIndexk}$, in \eqref{eq:GeneralizedMeasurementsVec} for any $r \in \mathbb{N}$, by taking weighted linear combinations \eqref{eq:WeightedSummationGeneralizedMeasurements} of the sensor data $\varphi_n(t_l)$ is equivalent to reproducing the function $\Psi_{\VecIndexk}(\Vecx) \Gamma_r(t)$ from space- and time-reversed translates of the Green's function $g(\Vecx,t)$ of the underlying field. Specifically, imposing $\widehat{\mathcal{Q}}(\VecIndexk,r) = {\mathcal{Q}}(\VecIndexk,r)$ implies that:
\begin{displaymath}
	\sum_{n=1}^{N} \sum_{l=0}^L w_{n,l}(\VecIndexk,r) g(\Vecx_n - \Vecx, t_l - t) \equiv \Psi_{\VecIndexk} (\Vecx) \Gamma_r(t).
\end{displaymath}
Furthermore when $\Psi_{\VecIndexk}(\Vecx)$ and $\Gamma_r(t)$ are chosen to be exponentials, this results in a multidimensional exponential reproduction problem.
\begin{IEEEproof}
We commence this proof by noting that \eqref{eq:Convolution_GreensFunc} can be written as:
\begin{align*}
u(\Vecx,t) &= f(\Vecx,t) \ast g(\Vecx,t) \\
					 &= \int_{\Vecx' \in \mathbb{R}^d}  \int_{t' \in \mathbb{R}} g(\Vecx', t') f(\Vecx - \Vecx', t - t') \,\mathrm{d}t' \mathrm{d}\Vecx' \\
					 &= \langle f(\Vecx', t'), g(\Vecx - \Vecx', t - t') \rangle_{\Vecx',t'},
\end{align*}
where $\Vecx' = (x'_1,x'_2, \ldots, x'_d) \in \mathbb{R}^d$ and $\mathrm{d}\Vecx' = \prod_{i=1}^d \mathrm{d}x_i'$. Consequently, the discrete measurement obtained by the $n$-th sensor (located at $\Vecx_n$) at some time instant $t_l \geq 0$ is
\begin{equation}
\varphi_n(t_l) = u(\Vecx_n,t_l) = \langle f(\Vecx, t), g(\Vecx_n - \Vecx, t_l - t) \rangle_{\Vecx,t}.
\label{eq:SensorMeasurements_InnerProd}
\end{equation}
Replacing \eqref{eq:SensorMeasurements_InnerProd} into the left hand side (lhs) of \eqref{eq:WeightedSummationGeneralizedMeasurements} yields:
\begin{align}
&\sum_{n = 1}^N \sum_{l=0}^L {w_{n,l}(\VecIndexk,r) \varphi_{n} (t_l)} 	= \sum_{n = 1}^N \sum_{l=0}^L {w_{n,l}(\VecIndexk,r) u(\Vecx_n, t_l)} \nonumber \\
																									&= \sum_{n = 1}^N \sum_{l=0}^L w_{n,l}(\VecIndexk,r) \left\langle f(\Vecx, t), g(\Vecx_n - \Vecx, t_l - t) \right \rangle_{\Vecx,t} \nonumber \\
																									&= \left \langle f(\Vecx, t), \sum_{n = 1}^N  \sum_{l=0}^L w_{n,l}(\VecIndexk,r) g(\Vecx_n - \Vecx, t_l - t) \right \rangle_{\Vecx,t},
\label{eq:InnerProd_GreenFuncCoeffs}
\end{align}
where $\{w_{n,l}(\VecIndexk,r)\}_{n,l} \in \mathbb{C}$ denote the specific sequence of weights we wish to compute\footnote{In the last equality, we are able to pass the summation inside the inner product because it is finite. If it were infinite then we would require that the sum converges absolutely; which is ensured if $g$ and its translates form a Riesz basis.}. In particular if we require this weighted sum of the sensor data to yield the exact multidimensional measurements, i.e. $\widehat{\mathcal{Q}}(\VecIndexk,r) = \mathcal{Q}(\VecIndexk,r) = \left\langle f(\Vecx,t), \Psi_{\VecIndexk}(\Vecx)\Gamma_r(t)  \right\rangle_{\Vecx,t}$, 
then by comparing the inner products in \eqref{eq:GeneralizedMeasurementsVec} and \eqref{eq:InnerProd_GreenFuncCoeffs}, we realise that we must choose the sequence of weights $\{w_{n,l}(\VecIndexk,r)\}_{n,l}$ such that, for each $\VecIndexk$ and $r$, the identity
\begin{equation}
\sum_{n = 1}^N \sum_{l=0}^L w_{n,l}(\VecIndexk,r) g(\Vecx_n - \Vecx, t_l - t) \equiv \Psi_{\VecIndexk} (\Vecx) \Gamma_r(t)
\label{eq:SpatiotemporalSensingFuncReproducing}
\end{equation}
is satisfied. This proves the first claim of the proposition.

The second statement follows immediately. If we impose the choice of sensing functions of \Cref{sec:SensingFunctions}, wherein $\Psi_{\VecIndexk}(\Vecx) = e^{\imagj \VecIndexk \cdot \Vecx}$ and $\Gamma_r(t) = e^{\imagj r t/T}$, then \eqref{eq:SpatiotemporalSensingFuncReproducing} becomes
	\begin{equation}
		\sum_{n = 1}^N \sum_{l=0}^L  w_{n,l}(\VecIndexk,r) g(\Vecx_n - \Vecx, t_l - t)  = e^{\imagj \VecIndexk \cdot \Vecx} e^{\imagj r t/T}
	\label{eq:GreensFunctionExponentialReprodution_2D}
	\end{equation}	
which can be written in a more compact form as follows:
\begin{equation}
	\sum_{n, l} w_{n,l}(\VecIndexk,r) g(\Vecx_n - \Vecx, t_l - t) = e^{\imagj (\VecIndexk,r/T) \cdot (\Vecx,t)} = e^{ \Veckappa \cdot (\Vecx,t)},
\label{eq:SpaceTimeExpReprodProblem}
\end{equation}
where $\Veckappa = \imagj (\VecIndexk,r/T) \in \mathbb{R}^{d+1}$. Hence we can immediately observe that the required coefficients are those that reproduce the $(d+1)$-dimensional (space and time varying) exponentials by summing translates of $g(-\Vecx,-t)$, which is the space- and time-reversed Green's function of the underlying field $u(\Vecx,t)$.
\label{prop:WeightedSum2GenMeasurements}
\end{IEEEproof}
\end{myProp}
Hence with access to the desired coefficients, $\{ w_{n,l}(\VecIndexk,r)\}_{n,l}$, that are capable of reproducing exponentials from the translates of $g(\Vecx,t)$, all we would need to do is: evaluate the sequence $\{\mathcal{Q}(\VecIndexk,r)\}_{\VecIndexk}$ for a fixed $r \neq 0$ using \eqref{eq:WeightedSummationGeneralizedMeasurements} and then from $\{\mathcal{Q}(\VecIndexk,r)\}_{\VecIndexk}$, extract the unknown source parameters as described in \Cref{sec:SensingFunctions}. Therefore the only missing piece in our framework is how to obtain the exponential reproducing coefficients. For this, one needs to understand when the exponential reproduction problem is exactly or approximately feasible and then establish appropriate schemes to find the desired coefficients.

We address these questions by leveraging from results in generalised sampling and approximation theory. Then for both (exact and approximate reproduction) cases, we derive closed-form formulae to compute the ``best'' weights $w_{n,l}(\VecIndexk,r)$, when the translates of the approximant are assumed to be regular. In the sensor network setup this is equivalent to having uniform spatiotemporal samples with sampling intervals $\bm{\Delta}_{\Vecx} = \left( \Delta_{x_1}, \Delta_{x_2}, \ldots, \Delta_{x_d} \right)$ and $\Delta_t $. Finally in the nonuniform sampling case, where it is generally not possible to obtain simple closed-form expressions for the desired exponential reproducing coefficients $w_{n,l}(\VecIndexk,r)$, we propose two approaches:
\begin{enumerate}
	\item Formulating and solving the linear system that comes from discretizing \eqref{eq:SpaceTimeExpReprodProblem}.
	\item Interpolating and resampling the sensor data uniformly.
\end{enumerate}
\subsection{Function spaces, generalised sampling and function approximation}
In the generalised (uniform) sampling paradigm---see \cite{Aldroubi2001_Nonuniform,Unser2002_Sampling,Vetterli2014_Foundations} and references therein---the primary goal is to reconstruct some functions of a continuous variable from a discrete set of measurements collected on a uniform grid. Often, this reconstruction will be an approximation of the original signal in some function spaces with a further property that the approximation error decays to zero, in the limit as the ``density'' of the sampling grid increases.

Consider the $d$-dimensional generating function $p$, whose uniform translates generates the space $$V_{\bm{\Delta}_{\Vecx}}(p) = \SPAN_{\VecIndexn \in \mathbb{Z}^d} \left\{ p\left( {\Vecx}/{\bm{\Delta}_{\Vecx}} - \VecIndexn \right) \right\}\text{,}$$ then any function $\hat{h}(\Vecx) \in V_{\bm{\Delta}_{\Vecx}}(p) \subset \Elltwo$ is characterised by the sequence of coefficients $a_{\VecIndexn}$, such that:
\begin{equation}
	\tilde{h}(\Vecx) = \sum_{\VecIndexn \in \mathbb{Z}^d} a_{\VecIndexn} p\left( {\Vecx}/{\bm{\Delta}_{\Vecx}} - \VecIndexn \right),
	\label{eq:GeneralizedSampling}
\end{equation}
where $\Elltwo$ denotes the space of square-integrable functions. In fact for this series to be well posed we require that:
\begin{enumerate}
	\item For convergence, $\left\{a_{\VecIndexn}\right\}_{\VecIndexn}$ must be square-summable.
	\item For uniqueness and stability of this discrete representation, $\left\{p(\Vecx - \VecIndexn)\right\}_{\VecIndexn}$ must form a Riesz basis of $V_{\bm{1}}(p)$.
	\item Finally, $p(\Vecx)$ must satisfy the partition of unity condition
		\begin{equation}
			\sum_{\VecIndexn \in \mathbb{Z}} p(\Vecx + \VecIndexn) = 1,
			\label{eq:PartitionOfUnity}
		\end{equation}
	for all $\Vecx \in \mathbb{R}^d$ in order to guarantee that by choosing $\bm{\Delta}_{\Vecx}$ in \eqref{eq:GeneralizedSampling} sufficiently small, we can approximate any function $\tilde{h}(\Vecx)$ as closely as we want (a detailed proof of this fact can be found in \cite[Appendix~B]{Unser2002_Sampling}).
\end{enumerate}
If we instead want to reconstruct some signal $h(\Vecx) \in \Elltwo \setminus \! V_{\bm{\Delta}_\Vecx}$, then the reconstruction $\tilde{h}(\Vecx)$ in \eqref{eq:GeneralizedSampling} should produce the best approximation of $h(\Vecx)$ in the space $ V_{\bm{\Delta}_\Vecx}$, and hence minimise the approximation error in the least-squares sense. This is achieved by computing the orthogonal projections of $h(\Vecx)$ onto $V_{\bm{\Delta}_{\Vecx}}(p)$ which is obtained by choosing $a_{\VecIndexn} = \left\langle h(\Vecx), p_{\text{dual}} \left({\Vecx}/{\bm{\Delta}_{\Vecx}} - \VecIndexn \right)\right\rangle$, where $p_{\text{dual}}$ is the dual of $p$ and is given by \cite{Blu1999_Quantitative},
\begin{equation*}
	\hat{p}_{\text{dual}}(\bm{\omega}) = \frac{\hat{p}(\bm{\omega})}{\sum_{\VecIndexn} \left\lvert \hat{p}(\bm{\omega} + 2\pi \VecIndexn) \right\rvert^2}.
\label{eq:DualBasisFourierDomain}
\end{equation*}
Here $\hat{p}(\bm{\omega}) = \mathcal{F}_{\Vecx} \{ p \} = \int_\Vecx p(\Vecx) e^{-\imagj \bm{\omega}\cdot\Vecx} \mathrm{d}\Vecx $ denotes the multidimensional Fourier transform of $p$. Please note also that in the rest of this paper, it is assumed that all transforms are taken in the sense of distributions.

Within our proposed framework we are seeking the specific coefficients $\{w_{n,l}(\VecIndexk,r)\}_{n,l}$ that reproduce the exponential function using shifted versions of the Green's function of the underlying physical field. Hence this is a special case of the above, where: 
\begin{enumerate}
	\item the signal we want to reconstruct is a specific $(d+1)$-dimensional exponential (i.e. $e^{\Veckappa \cdot (\Vecx,t)}$), and
	\item the generating function $p$ is precisely the Green's function of the underlying PDE.
\end{enumerate}
Under these conditions we want to find the best representation of the exponentials, in the space spanned by the translates of the Green's functions.
\subsection{Exact and approximate Strang-Fix theory for exponential reproduction from uniform translates}
\label{ssec:C6_ExactApproxStrangFix}
Building on the discussion of the previous section, we now focus on the approximation of exponentials from uniform translates of a single prototype function. We begin by proving the following lemma which is an extension of \cite{Strang1973_Fourier} using a multidimensional generalisation of the proof given in \cite{Uriguen2013FRI}.
\begin{myLem}[Generalised multidimensional Strang-Fix conditions \cite{Khalidov2005_Generalized}]
Let $p(\Vecx)$ be of compact support and its multidimensional bilateral Laplace transform be $P(\Vecs) = \int_{\Vecx \in \mathbb{R}^d}  g(\Vecx,t) e^{- \Vecx\cdot\Vecs}  \mathrm{d}\Vecx$, then the following conditions are equivalent:
\begin{enumerate}
	\item For any $ \VecIndexn \in \mathbb{Z}^d \setminus \! \{\bm{0}\}$, where $\bm{0}$ is the zero $d$-vector,
				\begin{equation}
				P(\Veckappa) \neq 0 \text{, whilst } P(\Veckappa + \imagj 2\pi \VecIndexn ) = 0.
				\label{eq:StrangFix_Condition}
				\end{equation}
	\item For some coefficients $ a_{\VecIndexn} \in \mathbb{C}$,
				\begin{equation}
					\sum_{\VecIndexn \in \mathbb{Z}^d} { a_{\VecIndexn} p(\Vecx -  \VecIndexn ) },
					\label{eq:StrangFix_PolyRep}
				\end{equation}
				is an exponential in $\Vecx$, i.e. $\sum_{\VecIndexn \in \mathbb{Z}^d} { a_{\VecIndexn} p(\Vecx -  \VecIndexn ) } = C e^{ \Veckappa \cdot \Vecx }$, for some $C \neq 0$.
\end{enumerate}
\label{lem:GenStrangFixConditions}
\begin{IEEEproof}
See Appendix \ref{app:C6_GenStrangFix}.
\end{IEEEproof}
\end{myLem}

The assumption of compactness here ensures first that the bilateral Laplace transform exists, so that the conditions \eqref{eq:StrangFix_Condition} are well-defined and second, that \eqref{eq:StrangFix_PolyRep} converges. This assumption is sufficient, but not necessary, since for example a suitable polynomial decay \cite{Light1992_Quasi}, or even milder restriction on $p(\Vecx)$ \cite{DeBoor1994_Approximation,Blu1999_Approximation} would still guarantee convergence. For the class of functions satisfying the generalised Strang-Fix conditions above, the desired coefficients can be computed exactly. 
To treat our multidimensional problem, we now extend formally the one-dimensional formulae obtained in \cite{Uriguen2013FRI} to the multidimensional case. This new formulae are still valid even when the prototype function is not separable with respect to its variables -- i.e we do not require that $p(\Vecx) = \prod_{i=1}^d{p_i(x_i)}$. The absence of this separability property is of paramount importance for us, especially because the spatial and temporal dimensions for most non-static fields encountered, in reality, are neither separable nor homogeneous. We begin our derivation by recalling that we are after the coefficients $\{a_{\VecIndexn}\}_{\VecIndexn}$ such that 
\begin{equation}
	\sum_{\VecIndexn \in \mathbb{Z}^d} a_{\VecIndexn} p(\Vecx - \VecIndexn) = e^{\Veckappa \cdot \Vecx}.
	\label{eq:ExpReprodProblem1}
\end{equation}

According to generalised sampling theory, the sequence of weights that minimises the approximation error in the least-squares sense is given by
\begin{align}
	a_{\VecIndexn}&= \left \langle e^{\Veckappa \cdot \Vecx}, p_{\text{dual}}(\Vecx - \VecIndexn) \right \rangle_{\Vecx} = \int_{\Vecx \in \mathbb{R}^d} e^{\Veckappa \cdot \Vecx} p_{\text{dual}}(\Vecx - \VecIndexn) \mathrm{d}\Vecx \nonumber \\
								&= \int_{\Vecx' \in \mathbb{R}^d} e^{\Veckappa \cdot (\Vecx' + \VecIndexn)} p_{\text{dual}}(\Vecx') \mathrm{d}\Vecx' \nonumber \\
								&= e^{\Veckappa \cdot \VecIndexn} \int_{\Vecx' \in \mathbb{R}^d} e^{\Veckappa \cdot \Vecx'} p_{\text{dual}}(\Vecx') \mathrm{d}\Vecx' \nonumber \\
								&= e^{\Veckappa \cdot \VecIndexn} a_{\bm{0}},
\label{eq:ExpReproCoeffs_an}
\end{align}
where the second line follows from the change of variable $\Vecx' = \Vecx - \VecIndexn$.
Thus finding $a_{\bm{0}}$ allows us to compute $a_{\VecIndexn}$ for all $\VecIndexn \in \mathbb{Z}^d$ using \eqref{eq:ExpReproCoeffs_an}. To find $a_{\bm{0}}$ substitute \eqref{eq:ExpReproCoeffs_an} into \eqref{eq:ExpReprodProblem1} to get 
$$ a_{\bm{0}}\!\! \sum_{\VecIndexn \in \mathbb{Z}^d} \! e^{\Veckappa \cdot \VecIndexn} p(\Vecx - \VecIndexn) {=} e^{\Veckappa \cdot \Vecx}  \Leftrightarrow a_{\bm{0}}\!\! \sum_{\VecIndexn \in \mathbb{Z}^d} \! e^{-\Veckappa \cdot (\Vecx - \VecIndexn )} p(\Vecx - \VecIndexn) {=} 1.$$
We then apply Poisson summation formula on the lattice to the l.h.s. of this expression, which if $p(\Vecx)$ is well-behaved reduces to 
$$a_{\bm{0}}\sum_{\VecIndexn \in \mathbb{Z}^d} P(\Veckappa + \imagj 2\pi \VecIndexn ) e^{\imagj 2 \pi \VecIndexn \cdot \Vecx} = 1.$$
Finally from the condition \eqref{eq:StrangFix_Condition} we get $a_{\bm{0}} = \frac{1}{P(\Veckappa)}$, since all terms in the summation vanish for $\VecIndexn \neq \bm{0}$. Hence for any $\VecIndexn \in \mathbb{Z}^d$ it follows that,
\begin{equation}
	 a_{\VecIndexn} = \frac{e^{\Veckappa \cdot \VecIndexn}}{P( \Veckappa)}.
\label{eq:ExpRepCoeffs_Exact}
\end{equation}
\subsubsection{Approximate Strang-Fix in multidimensions and the approximation error}
In the derivation of \eqref{eq:ExpRepCoeffs_Exact} we imposed some regularity conditions on $p(\Vecx)$, specifically for the l.h.s of the Poisson summation formula to converge, the function must decay sufficiently quickly. The strongest constraint on $p(\Vecx)$ however is due to \eqref{eq:StrangFix_Condition}, where $P(\Veckappa + \imagj 2\pi \VecIndexn ) = 0$ for $\VecIndexn \in \mathbb{Z}^d \setminus \! \{\bm{0}\}$.

For general physical fields of interest to us, the approximant $p(\Vecx)$ will be replaced by the corresponding Green's function $g$ of the field. Whilst these will generally not satisfy the Strang-Fix condition \eqref{eq:StrangFix_Condition}, we still wish to reproduce exponentials approximately with them. To this end, we will extend the approximate Strang-Fix method introduced in \cite{Uriguen2013FRI}, which relaxes the assumptions on the generators (for the $1$-D exponential case), so that we are now after the coefficients that gives the best approximate exponential reproduction, for any $p(\Vecx)$. Mathematically this means that we desire
\begin{equation}
\sum_{\VecIndexn \in \mathbb{Z}^d} a_{\VecIndexn} p(\Vecx-\VecIndexn) \approx e^{\Veckappa \cdot \Vecx},
\label{eq:ExponentialReproduction_Approx}
\end{equation}
where $p(\Vecx)$ does not necessarily satisfy the generalised Strang-Fix conditions \eqref{eq:StrangFix_Condition}. There are a few possible choices one may make for the ``best'' approximation coefficients. For any choice, the associated approximation error is, $\varepsilon(\Vecx)  = e^{\Veckappa \cdot \Vecx}\left(1 - a_{\bm{0}} \sum_{\VecIndexn} P(\Veckappa + \imagj 2\pi \VecIndexn ) e ^{\imagj 2 \pi \VecIndexn \cdot \Vecx} \right)$ which can be minimised in the least-squares sense, by computing the orthogonal projection of $e^{\Veckappa \cdot \Vecx}$ on the the subspace $V_{\bm{1}}(p)$. This yields $a_{\bm{0}} = \frac{P(-\Veckappa)}{R_p(e^{\Veckappa})}$, where $R_p(e^{\Veckappa}) = \sum_{\bm{\ell} \in \mathbb{Z}^d} r_p[\bm{\ell}] e^{-\Veckappa \cdot \bm{\ell}}$ is the multidimensional $z$-transform of the autocorrelation sequence $r_p[\bm{\ell}] = \left\langle p(\Vecx - \bm{\ell}),p(\Vecx)\right\rangle_{\Vecx}$, evaluated at $\mathbf{z} = e^{\Veckappa} = (e^{\kappa_1},e^{\kappa_2},\ldots,e^{\kappa_d})$ \cite{Uriguen2013FRI,Blu1999_Approximation}.

Moreover, observe that the square error $\varepsilon^2(\Vecx)$ is minimised when $ 1 - a_{\bm{0}} \sum_{\VecIndexn} P(\Veckappa + \imagj 2\pi \VecIndexn ) e ^{\imagj 2 \pi \VecIndexn \cdot \Vecx} = 0$; in addition if the Laplace transform $P$ of the generator decays quickly, i.e. assuming $P(\Veckappa + \imagj 2\pi \VecIndexn ) \approx 0$ for any $\VecIndexn \in \mathbb{Z}^d\setminus \{\bm{0}\}$, then $a_{\bm{0}} = 1/P(\Veckappa)$ is a good proxy for the minimiser of $\varepsilon(\Vecx)$ and therefore:
\begin{equation}
	a_{\VecIndexn}(\Veckappa) = \frac{e^{\Veckappa \cdot \VecIndexn}}{P(\Veckappa)}.
\label{eq:ConstantLeastSquaresCoeffs}
\end{equation}

These are the coefficients we shall utilise in this work, for their simplicity and accuracy. Moreover, the approximation error using \eqref{eq:ConstantLeastSquaresCoeffs} is given by:
\begin{equation}
	\varepsilon(\Vecx)  = e^{\Veckappa \cdot \Vecx}\left(1 - \frac{1}{P(\Veckappa)} \sum_{\VecIndexn} P(\Veckappa + \imagj 2\pi \VecIndexn ) e^{\imagj 2 \pi \VecIndexn \cdot \Vecx} \right),
	\label{eq:ConstantLeastSquaresCoeffs_Error}
\end{equation}
which will be small if $P(\Veckappa + \imagj 2\pi \VecIndexn )$ decays quickly to zero as $|\VecIndexn|$ increases.

\subsection{Computing the analysis coefficients for space-time fields}
\label{ssec:ExplicitCoefficients}
Equipped with the formulae in \Cref{ssec:C6_ExactApproxStrangFix}, we can now compute explicitly the weights $w_{n,l}(\VecIndexk,r)$ in \eqref{eq:WeightedSummationGeneralizedMeasurements}. We distinguish two cases: \begin{inparaenum}[\itshape(i)] \item the case where the sensors are uniformly spaced and \item the case where the location of the sensors are arbitrary.\end{inparaenum}
\subsubsection{Uniform sensor placement}
Denote the sensor measurements by $\varphi_{\VecIndexn}(t_l) = u(\VecIndexn \bm{\Delta}_{\Vecx}, l \Delta_t)$, where $\VecIndexn \bm{\Delta}_{\Vecx} = (n_1\Delta_{x_1}, n_2\Delta_{x_2}, \ldots, n_d\Delta_{x_d})$ and $n_i = 0,1,\ldots, N_i - 1$ for $i = 1,\ldots, d$. Note that we can reconcile the vector index sensor measurement $\varphi_{\VecIndexn}(t_l)$ with the scalar indexed one $\varphi_{n}(t_l)$, by simply taking the lexicographic ordering of the elements of $\{ \VecIndexn \Delta_{\Vecx} \}_{\VecIndexn \in \mathbb{N}^d}$ to give $\{\Vecx_n\}_{n=1}^N$, where $N = \prod_{i=1}^d N_i$.

\begin{figure*}[tb]
\centering
\subfloat[$\VecIndexk = (2,2)$.]{\includegraphics[width=0.33\linewidth,trim=0 0.8cm 0 1cm, clip]{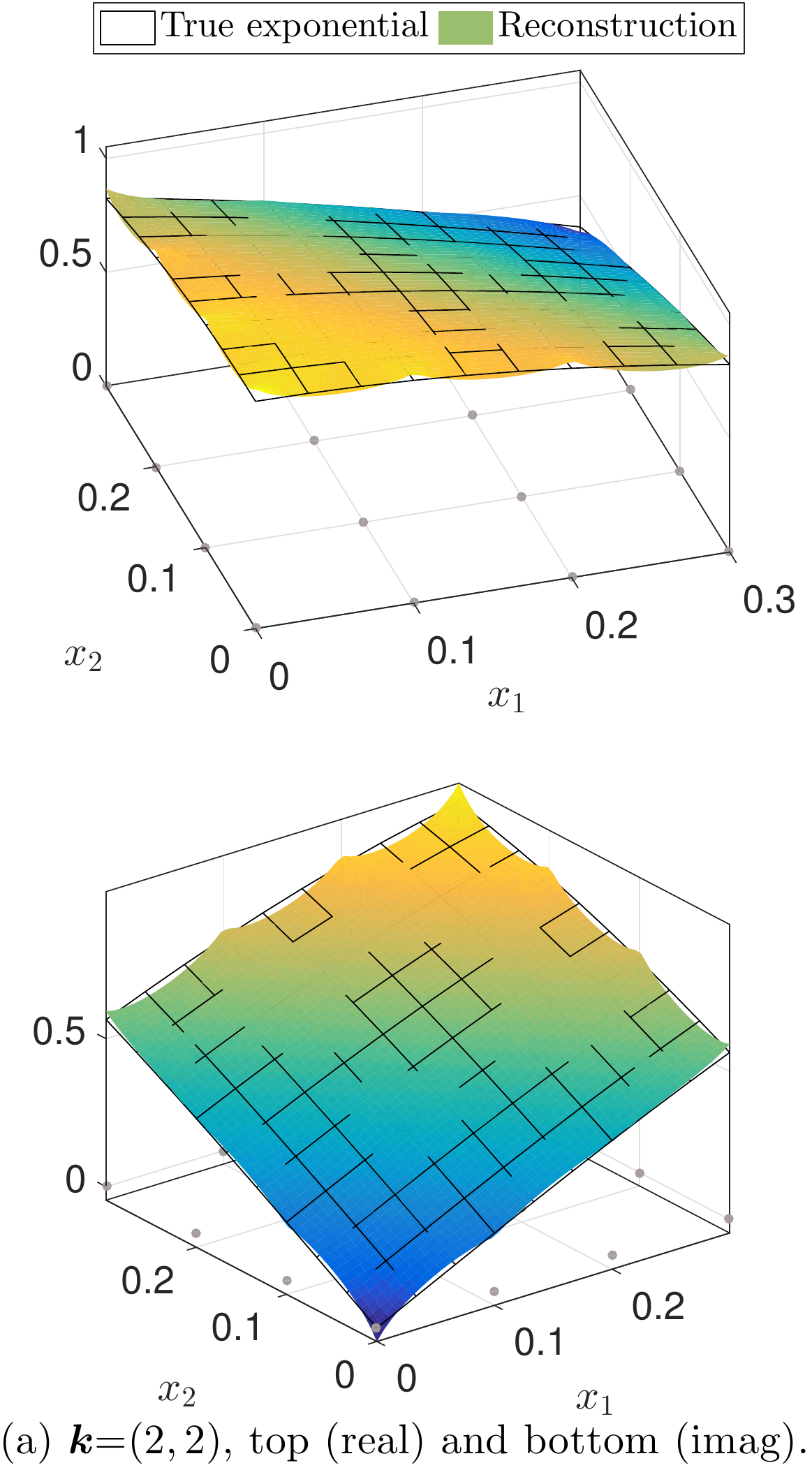}}
\subfloat[$\VecIndexk = (2,3)$.]{\includegraphics[width=0.32\linewidth,trim=0 0.9cm 0 0cm, clip]{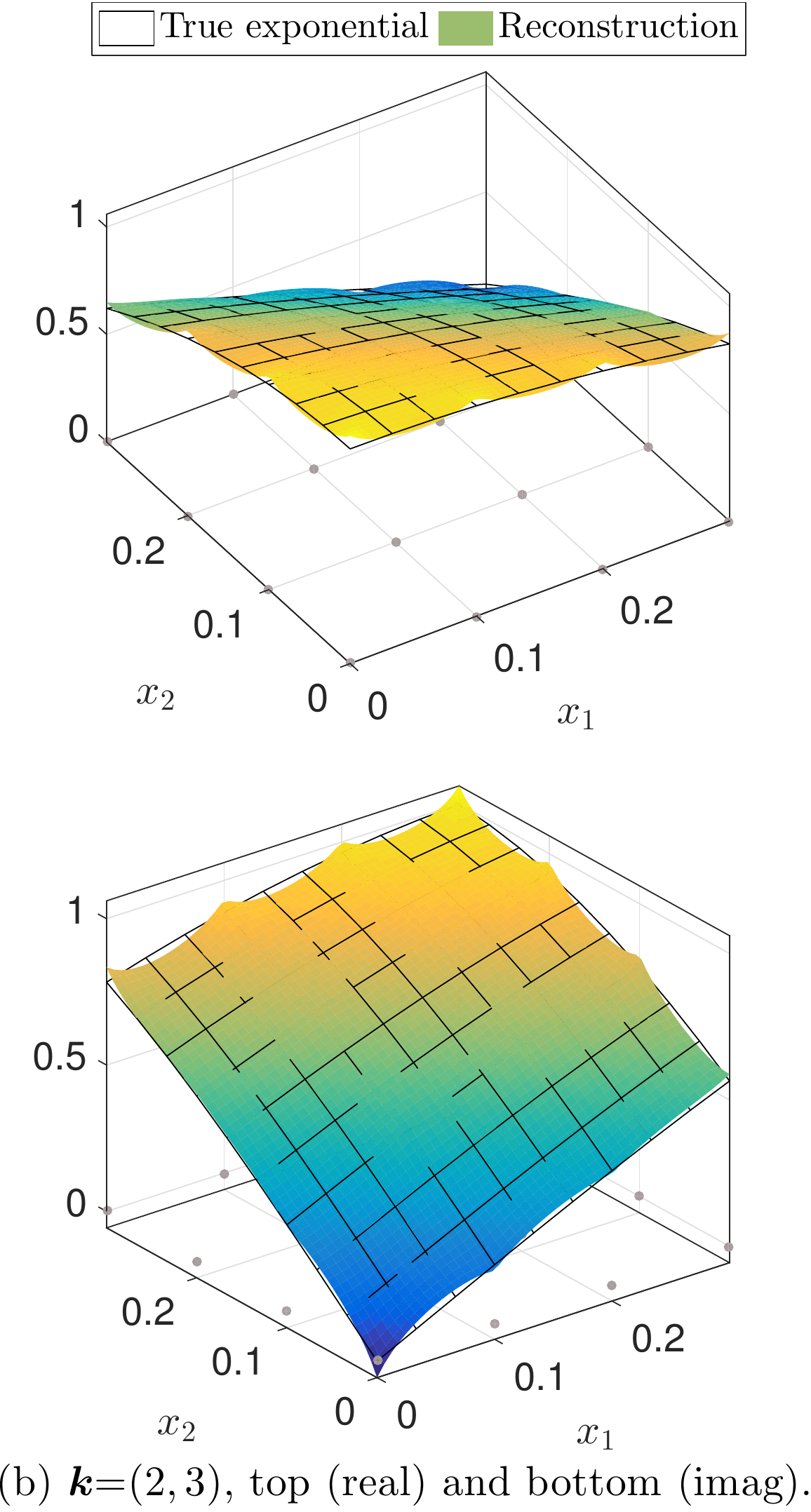}}
\subfloat[$\VecIndexk = (2,4)$.]{\includegraphics[width=0.31\linewidth,trim=0 0.85cm 0 1cm, clip]{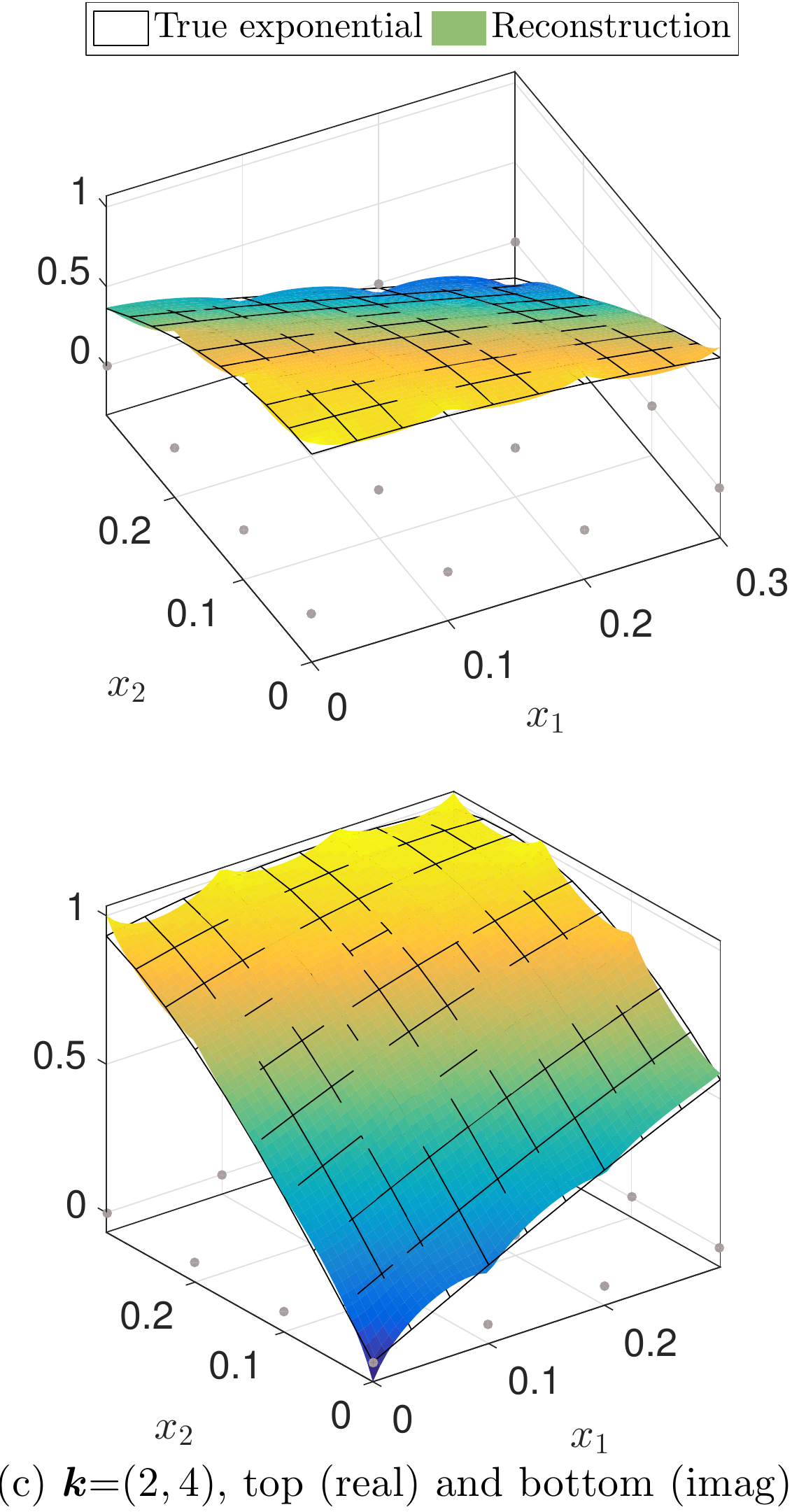}}
\caption{\textbf{Exponential reproduction. }Reproducing the $2$-D sensing function $\Psi_{\VecIndexk}(\Vecx)=e^{\imagj k_1 x_1 + \imagj k_2 x_2}$, assuming $N = 16$ uniformly placed sensors (grey `$\bullet$') for the $2$-D potential field. Specifically the translates of the prototype \eqref{eq:GreensFunction_Poisson2D}, centred at the indicated grey `$\bullet$' locations, are linearly combined using the exponential reproducing weights obtained with \eqref{eq:ExpReproducingCoeffs_NonInteger}. This gives the approximate Strang-Fix reconstruction (solid surface) of the exponentials (black mesh).}
\label{fig:ApproximationOfSTSF}
\end{figure*}

Consequently for physics-driven fields, it is clear from \eqref{eq:SpaceTimeExpReprodProblem} that the prototype function is the space- and time-reversed Green's function: $p = g(-\Vecx, -t)$, with bilateral Laplace transform $G(-\Vecs_{\Vecx}, -s_t)$. Therefore for translates $\bm{\Delta}_{\Vecx} \in \mathbb{R}_{+}^{d}$ and $\Delta_t \in \mathbb{R}_{+}$ the corresponding exponential reproducing coefficients are:
\begin{equation}
\bar{w}_{\VecIndexn,l}(\VecIndexk, r) = \frac{ e^{\bar{\Veckappa}\cdot(\VecIndexn,l)}}{\bar{G}(- \imagj \bm{\Delta}_{\Vecx} \VecIndexk, - \Delta_t \imagj r/T)},
\label{eq:ExpReprodCoeffs_gbar}
\end{equation}
where $\bar{g}(\Vecx,t) = g(\bm{\Delta}_{\Vecx}\Vecx, \Delta_t t )$ and $\bar{\Veckappa} = \imagj(\bm{\Delta}_{\Vecx} \VecIndexk, \Delta_t r/T)$. Since, $\bar{g}(\Vecx,t) \Leftrightarrow \bar{G}(\Vecs_{\Vecx},s_t) = \frac{1}{\Delta_t \prod_{i=1}^d \Delta_{x_i}} G(\frac{\Vecs_{\Vecx}}{\bm{\Delta}_{\Vecx}},\frac{s_t}{\Delta_t})$, we can find $\bar{G}(- \imagj \bm{\Delta}_{\Vecx} \VecIndexk, - \Delta_t \imagj r/T)$ and substitute it into \eqref{eq:ExpReprodCoeffs_gbar}, to conclude that for any $\bm{\Delta}_{\Vecx} \in \mathbb{R}^d_{+}$ and $\Delta_t \in \mathbb{R}_{+}$,
\begin{equation}
w_{\VecIndexn,l}(\VecIndexk, r) = \Delta_t \prod_{i=1}^d \Delta_{x_i} \frac{e^{\imagj( \bm{\Delta}_{\Vecx}\VecIndexk,\Delta_t r/T)\cdot(\VecIndexn,l)}}{{G}(- \imagj \VecIndexk, -\imagj r/T)}.
\label{eq:ExpReproducingCoeffs_NonInteger}
\end{equation}
As an example, we show in \Cref{fig:ApproximationOfSTSF}, the approximation of the 2-D spatial exponentials, $e^{\imagj k_1 x_1 + \imagj k_2 x_2}$, using the 2-D Green's function of Poisson's equation \eqref{eq:GreensFunction_Poisson2D} and the coefficients \eqref{eq:ExpReproducingCoeffs_NonInteger}, for $r=0$, $k_1 = 2$, and $k_2 = 2, 3, 4$.

\subsubsection{Least-squares scheme for nonuniformly placed sensors}
\label{sssec:MatrixInversion}
For the case of non-uniformly placed sensors, it is generally not possible to find similar closed-form expressions for the desired coefficients $\{w_{n,l}(\VecIndexk,r)\}_{n,l}$. It is possible however to formulate a linear system of equations to find $\{w_{n,l}(\VecIndexk,r)\}_{n,l}$ since the approximating function $g$ and the exponentials $\Gamma_r(t) \Psi_{\VecIndexk}(\Vecx)$ we want to approximate are known. One approach is to discretise \eqref{eq:SpatiotemporalSensingFuncReproducing} as follows. First, for each $l$, formulate the following linear system at some fixed time snapshot $t_j > 0$:\vspace{-3mm}%
\begin{equation*}
\begin{gathered}
\left[ \begin{array}{cccc}
	g(\Vecx_1 {-} \Vecx'_1, t_l {-} t_j)  & \cdots & g(\Vecx_N {-} \Vecx'_1, t_l {-} t_j) \\
	g(\Vecx_1 {-} \Vecx'_2, t_l {-} t_j)  & \cdots & g(\Vecx_N {-} \Vecx'_2, t_l {-} t_j) \\
					 \vdots							      & \ddots & 					\vdots								 \\
	g(\Vecx_1 {-} \Vecx'_I, t_l {-} t_j)  & \cdots & g(\Vecx_N {-} \Vecx'_I, t_l {-} t_j)
\end{array} \right] \! \!
\left[\arraycolsep=0.015pt\def\arraystretch{1.2} \begin{array}{c}
	w_{1,l}(\VecIndexk,r) \\
	w_{2,l}(\VecIndexk,r) \\
	\vdots \\
	w_{N,l}(\VecIndexk,r)
\end{array} \right] \\
 = \left[\arraycolsep=0.015pt\def\arraystretch{1.2} \begin{array}{c}
	\Psi_{\VecIndexk} (\Vecx'_1) \Gamma_r(t_j)\\
	\Psi_{\VecIndexk} (\Vecx'_2) \Gamma_r(t_j) \\
	\vdots \\
	\Psi_{\VecIndexk} (\Vecx'_I) \Gamma_r(t_j)
\end{array} \right]
\end{gathered}
\end{equation*}
\begin{align}
\Rightarrow \VecG_{l,j} \Vecw_{l}(\VecIndexk,r) &= \Vecp_{j}(\VecIndexk,r).
\label{eq:DirectMethod_LinearSys_Space}
\end{align}
Solving this system gives the coefficients $\{w_{n,l}(\VecIndexk,r)\}_{n}$ for a fixed $l$. Second, we stack \eqref{eq:DirectMethod_LinearSys_Space} for each $l$ and take several $t_j$'s, $j = 1,\ldots,J$ to get
\begin{align}
\left[ \arraycolsep=0.05pt\def\arraystretch{1} \begin{array}{cccc}
	\VecG_{0,1}  & \VecG_{1,1} & \cdots & \VecG_{L,1} \\
	\VecG_{0,2}  & \VecG_{1,2} & \cdots & \VecG_{L,2} \\
					 \vdots							    &				 &				\vdots								  \\
	\VecG_{0,J}  & \VecG_{1,J} & \cdots & \VecG_{L,J} \\
\end{array} \right]
\left[ \arraycolsep=0.05pt\def\arraystretch{1} \begin{array}{c}
	\Vecw_{0}(\VecIndexk,r) \\
	\Vecw_{1}(\VecIndexk,r) \\
	\vdots \\
	\Vecw_{L}(\VecIndexk,r)
\end{array} \right]
 &= \left[ \arraycolsep=0.05pt\def\arraystretch{1} \begin{array}{c}
	\Vecp_{1}(\VecIndexk,r)\\
	\Vecp_{2}(\VecIndexk,r) \\
	\vdots \\
	\Vecp_{J}(\VecIndexk,r)
\end{array} \right] \nonumber \\
\VecG \Vecw(\VecIndexk,r) &= \Vecp(\VecIndexk,r),
\label{eq:DirectMethod_LinearSystem_final}
\end{align}
where $\VecG \in \mathbb{R}^{IJ \times N(L+1)}$ is a discretisation of $g(\Vecx,t)$, $\Vecp(\VecIndexk,r) \in \mathbb{R}^{IJ}$ are discretisations of the spatiotemporal sensing functions, whilst $\Vecw(\VecIndexk,r) \in \mathbb{R}^{N(L+1)}$ are the desired weights for each $\VecIndexk $ and $r$.
Consequently, in order to recover the desired field analysis coefficients, we would need to solve the system \eqref{eq:DirectMethod_LinearSystem_final}. In general, this system admits a least-squares solution if $IJ \geq N(L+1)$, where the observation matrix $\VecG$ can be constructed from the Green's function of the problem at hand (i.e. \eqref{eq:GreensFunction_Poisson2D}, \eqref{eq:GreensFunction_Poisson3D}, \eqref{eq:GreensFunction_Diff}, \eqref{eq:GreensFunction_Wave2D} and so on).

Although straightforward to formulate, the conditioning of such a system can be poor in some instances. Specifically, the condition number of $\VecG$ depends directly on the sensor locations $\Vecx_n$, the sampling instants $t_l$ and the Green's function $g(\Vecx,t)$ of the underlying phenomena.%
\subsubsection{Interpolation scheme for nonuniform sampling}
\label{sssec:Interpolation}
Another simple, yet effective scheme for handling non-uniform sensor placement is to interpolate the field and resample it on a uniform grid. In so doing, we can obtain an approach that can still exploit the closed form expression \eqref{eq:ExpReproducingCoeffs_NonInteger} even when the spatial sampling is irregular.

Essentially we want to return to the situation where translates of the Green's function are on a uniform lattice. To do this, we assume that the spatial field samples are interpolated, on a uniform grid, using the interpolator family $\{\gamma_n(\Vecx)\}$, such that $\hat{u}(\Vecx,t_l) = \sum_{n=1}^N \varphi_n(t_l) \gamma_n(\Vecx - \Vecx_n)$. Then this new approximation of the underlying field is resampled uniformly at the new locations $\{ \Vecx_{\bar{\VecIndexn}} = (\bar{n}_1 \Delta_{x_1},\bar{n}_2 \Delta_{x_2},\ldots,\bar{n}_d \Delta_{x_d}) \}_{\bar{\VecIndexn} \in \mathbb{N}^d}$ to give the corresponding data samples $\hat{\varphi}_{\bar{\VecIndexn}}(t_l) = \hat{u}(\Vecx_{\bar{\VecIndexn}},t_l)$. Finally, since we are back to the uniform case, the corresponding exponential reproducing coefficients can be recovered using \eqref{eq:ExpReproducingCoeffs_NonInteger}. Then weighting the interpolated measurements $\hat{\varphi}_{\bar{\VecIndexn}}(t_l)$ by the obtained coefficients will produce an estimate for the desired sequence of generalised measurements. Besides avoiding the inversion of poorly-conditioned matrices, the interpolation approach is less intensive computationally compared to matrix inversion, particularly when the matrix $\VecG$ (in \eqref{eq:DirectMethod_LinearSystem_final}) is large. In our experimental results, we have used linear splines as the interpolator family $\{\gamma_n(\Vecx)\}$.
%
%

%
%
%
\section{The Sensor Network Algorithms: Source estimation from field samples}
\label{sec:SNAlgorithms}
Based on the framework outlined in \Cref{sec:SensingFunctions,sec:SensorData2GeneralizedMeasurements}, we now develop practical sensor network algorithms for estimating the sources of a physical field from its sensor measurements, and therefore solve the class of ISPs driven by linear PDEs with constant coefficients. First, we outline a centralised algorithm, which follows straightforwardly from \Cref{prop:WeightedSum2GenMeasurements}, then we consider the distributed case which relies on average consensus algorithms. %

\subsection{Centralised source estimation}
\label{ssec:CentralisedEstimation}
Assuming the source measurements have been made available at the fusion centre and that the PDE model---and Green's function---of the monitored phenomena are known, then the point and instantaneous source estimation scheme can be summarised as in \Cref{alg:UniversalSimultaneousPointSourceEst_Central}, when the number of field sources $M$ is known.
\begin{rem}
Notice that \Cref{alg:UniversalSimultaneousPointSourceEst_Central} requires $M$ as input, however when $M$ is not known, the following interesting phenomenon persists. Let ${M'}$ be our initial guess for the unknown $M$, if it is incorrect (i.e. when ${M'} \neq M$), we observe even in noisy settings, that:
	\begin{enumerate}
		\item If $ M' < M$, the most dominant $M'$ sources are recovered from the field measurements.
		\item Otherwise if $ M' > M$, all $M$ sources are recovered. In addition to them however, $ M' - M$ spurious sources---which may be attributed to noise and other sources of model mismatch---are also estimated. These extra $ M' - M$ spurious sources will either fall outside of the monitored region $\Omega$ or their estimated intensities will be very small in comparison to the true sources.
	\end{enumerate}
As we will see in \Cref{ssec:NoiseModelMismatch}, this observation can be conveniently exploited to develop a suitable estimation scheme even when $M$ is not known \textit{a priori}.
	\label{rem:UnknownNumSources}
\end{rem}%
%
%
%
\begin{algorithm}[hbt]
\caption{Simultaneous estimation of $M$ point sources}
\label{alg:UniversalSimultaneousPointSourceEst_Central}
\begin{algorithmic}[1]
\Require $\{ \varphi_n(t_l) \}_{n=1,l=0}^{N,L}$, $\{\Vecx_n\}_n$, $M$, $\Delta_t$, $\mu$, $d$.
\State Compute the Laplace transform $G(\Vecs_{\Vecx}, s_t)$ of the Green's function.
\State Initialise $K_i \geq 2M-1$ for each $i = 1, \ldots, d$ and $r=1$.
\If{\texttt{UniformSampling}}
	\State From $\{ \Vecx_n\}_n$ compute sensor spacing $\bm{\Delta}_{\Vecx}$.
	\State Compute coefficients $\{w_{\VecIndexn,l}(\VecIndexk,1)\}_{\VecIndexk=(0,0)}^{(K_1,K_2)}$ using \eqref{eq:ExpReproducingCoeffs_NonInteger}.	
\Else
	\State Use approach in \Cref{sssec:MatrixInversion} or \Cref{sssec:Interpolation}.
\EndIf
	\State Compute $\{\mathcal{Q}(\VecIndexk,1) = \sum_{\VecIndexn,l} w_{\VecIndexn,l}(\VecIndexk,1) {\varphi}_{\VecIndexn}(t_l) \}_{\VecIndexk}$.
	\State Recover all $M$ pairs of $(c_m e^{-\imagj \tau_m / T}, \bm{\xi}_m)$ by applying $N$-D ESPRIT (Appendix \ref{app:ParamEstimation_SoS}) to $\{\mathcal{Q}(\VecIndexk,1)\}_{\VecIndexk}$.
	\State For all $m$, $c_m {=} \left| c_m e^{-\imagj \tau_m / T} \right|$ and $ \tau_m {=} T \arg\mathopen{}\left(\! c_m e^{-\imagj \tau_m / T}\!\right)\mathclose{}$.
	\State \Return $\{c_m, \tau_m, \bm{\xi}_m\}_{m=1}^M$.
\end{algorithmic}
\end{algorithm}
%
%
\subsection{Distributed source estimation}
\label{ssec:Distr_SNalgorithm}
In the distributed set up, we want each node in the network to first estimate $\mathcal{Q}(\VecIndexk,r)$ through \textit{localised interactions} with its neighbouring nodes. These {localised interactions} in our field estimation setting are based on the use of consensus algorithms.

We assume the same sensor network as described in \Cref{ssec:SensorNetworkAssumptions}, comprising of ``smart'' sensor nodes. In addition to knowing the Green's function of the monitored phenomena\footnote{This comes for free since the SNs are designed to sense a particular phenomena: i.e. if we are sensing acoustic fields then we use the wave equation, for temperature and leakages we use the diffusion equation, and so on.}, these nodes are able to perform mathematical computations and can also learn the network topology upon deployment. Knowledge of the network topology and the Green's function of the underlying phenomena, means that all sensors can compute independently their sets of exponential reproducing coefficients $\{w_{n,l}(\VecIndexk,r)\}_l$ by using, either \eqref{eq:ExpReproducingCoeffs_NonInteger} if they are on a uniform grid or the least-squares scheme described in \Cref{sssec:MatrixInversion} when they are nonuniformly placed.

They may start to sense the field in time to measure $\varphi_{n}(t_l)$. To estimate the unknown sources the sensors must exchange and aggregate their sensor data, using average consensus as described in Appendix \ref{app:GossipAlgorithms}. Specifically, gossiping is initiated when the $n$-th sensor contacts and exchanges its local measure,
\begin{equation}
	y_{n}(\VecIndexk,r) = N \sum_{l=0}^L w_{n,l}(\VecIndexk,r) \varphi_{n}(t_l),
\label{eq:GossipLocalMeasures}
\end{equation}
with a neighbour. After several rounds of gossip, each sensor will converge to the generalised measurements,
\begin{equation*}
\frac{1}{N} \sum_{n=1}^N y_{n}(\VecIndexk,r) = \frac{1}{N} \sum_{n=1}^N N \sum_{l=0}^L w_{n,l}(\VecIndexk,r) \varphi_{n}(t_l) = \mathcal{Q}(\VecIndexk,r).
\end{equation*}
Consequently $N$-D ESPRIT, or a similar Prony-like method, is then applied by each sensor---independently---on $\{\mathcal{Q}(\VecIndexk,r)\}_{\VecIndexk}$, to compute locally the unknown source parameters.
\subsection{Estimation in presence of noise and model mismatch}
\label{ssec:NoiseModelMismatch}
Although the sensors actually acquire noisy measurements, i.e. $\{\varphi_{n,l}^{\epsilon}\}_{n,l}$ in \eqref{eq:NoisySamples}, of the underlying field the same recovery schemes outlined in \Cref{ssec:CentralisedEstimation,ssec:Distr_SNalgorithm} are still effective. Specifically mapping the noisy sensor samples, $\{\varphi_{n,l}^{\epsilon}\}_{n,l}$, to generalised measurements using \eqref{eq:WeightedSummationGeneralizedMeasurements} gives the noisy sequence,\vspace{-2mm}
\begin{equation*}\vspace{-3mm}
	\mathcal{Q}_{\epsilon}(\VecIndexk,r) = \sum_{n,l} w_{n,l}(\VecIndexk,r) \varphi_{n,l}^{\epsilon} = \mathcal{Q}(\VecIndexk,r) + \overbrace{\sum_{n,l} w_{n,l}(\VecIndexk,r) \epsilon_{n,l}}^{\text{coloured noise}}.
\end{equation*}
Thus to recover the field sources from $\{\mathcal{Q}_{\epsilon}(\VecIndexk,r)\}_{\VecIndexk=(0,\ldots,0)}^{(K_1,\ldots,K_d)}$, using $N$-D ESPRIT, we construct a (noisy) multilevel Hankel matrix $\VecH_{\epsilon}$ according to \eqref{eq:MultiLevelHankel}. If $M$ is known, then choosing $K_i \geq 2M - 1$ for all $i=1,\ldots,d$, and retaining only those singular vectors due to the $M$ largest singular values of $\VecH_{\epsilon}$ is implicitly denoising. Generally, choosing large $K_i$ for all $i$ promotes robustness even in low SNR regimes.

Often however, the multilevel Hankel structure of $\VecH_{\epsilon}$ is lost when we retain only its $M$ largest singular values (and zero the rest). However we may restore its structure by averaging the appropriate elements. These steps can be repeated until convergence. This is the fundamental idea behind Cadzow-like \cite{Cadzow1988_Signal} algorithms common in the FRI literature. Furthermore, these approaches are most effective when the noise in $\VecH_{\epsilon}$ is white, so we first need to apply a noise prewhitening transform similar to the approach in \cite{Murray1602_Physics}.
\subsubsection{Recovering an unknown number of sources}
\label{sssec:UnknownNumSources}
By combining the observations in \Cref{rem:UnknownNumSources} with the fact that the singular value decomposition of the multilevel Hankel matrix $\VecH$ (or $\VecH_{\epsilon}$) also encodes information about $M$---i.e. most dominant singular values are due to the sources---an iterative estimation scheme similar to those in \cite{Murray1506_Estimating,Murray1602_Physics} can be devised. The scheme relies on finding a time interval over which only a fixed number of field sources are active. Finding such a window allows us to reliably estimate these active sources and then adjust the spatiotemporal sensor measurements by removing their contribution to the sensor measurements. Given $\{\varphi_n(t_l)\}_{l=0}^L$, the strategy (to find a window with one active source) is as follows:
\begin{enumerate}
		\item Assume that there are $M'\geq 2$ sources and approximate $\{\mathcal{Q}(\VecIndexk,0)\}_{\VecIndexk}$ where $k_i = 0, 1, \ldots, 2M' - 1$, using only the samples $\{\varphi_n(t_l)\}_{l=0}^{L'}$ with $L' < L$ from the time window $[0,L' \Delta_t]$. The centralised or distributed approaches (in Section V) can be used.
		\item Proceed to estimate the $M'$ source intensities $\{c'_{m'}\}_{m'=1}^{M'}$ and locations $\{\bm{\xi}'_{m'}\}_{m'=1}^{M'}$ using the multidimensional ESPRIT method.
		\item Check that the estimated sources are valid. In particular, a source $(c'_{m'},\bm{\xi}'_{m'})$ is valid if the conditions below are simultaneously satisfied:
		\begin{enumerate}
			\item $c'_{m'}$ is greater than some predetermined threshold, and;
			\item $\bm{\xi}'_{m'}$ is within the monitored region.
		\end{enumerate}
		Check all $M'$ sources and let $M_{\text{vs}}$ be the number of valid sources found.
		\item There are three cases, if
		\begin{enumerate}[label= \roman*)]
			\item $M_{\text{vs}} > 1$: Reduce the time widow by reducing $L'$ and return to step (a).
			\item $M_{\text{vs}} < 1$: Increase the time widow by increasing $L'$ and return to step (a).
			\item $M_{\text{vs}} = 1$: Estimate the source parameters from the measurements $\{\varphi_n(t_l)\}_{l=0}^{L'}$ using \Cref{alg:UniversalSimultaneousPointSourceEst_Central}.
		\end{enumerate}
		\item Synthesise the field due to this source using equation (2) and adjust the sensor measurements by removing the contribution of this source. Increment $L'$ and return to step (a).
		\item Stop when the $L'=L$ or when the field measurements are below some predefined threshold.
\end{enumerate}

\subsection{Filtering in the time-domain}
\label{ssec:TemporalFiltering}
Using the framework summarised by \Cref{prop:WeightedSum2GenMeasurements}, we have been able to devise practical sensor network algorithms to solve the ISP of interest. During the sensing phase, although spatial prefiltering is generally not realisable, we are still able to perform prefiltering in time. The prefiltered samples obtained by the $n$-th sensor, using the filter $h(t)$ are: 
\begin{align*}
	\phi_n (t_l) &{=} \left.u(\Vecx_n, t) \star h(t) \right \rvert_{t {=} t_l} \\
							 &{=} \left.f(\Vecx,t) \ast g(\Vecx,t) \star h(t) \right \rvert_{\Vecx {=} \Vecx_n, t {=} t_l},
\end{align*}
where $\star$ is the time-convolution operator. In light of this new formulation, the generator that will be used to reproduce exponentials from its space-time translates is 
\begin{equation}
g_f(\Vecx,t) = \int_{t'} g(-\Vecx,-t-t') h(t') \mathrm{d}t',
\label{eq:PrefilteredGreensFunction}
\end{equation}
with $G_f(\Vecs_{\Vecx},s_t) = {G}(-\Vecs_{\Vecx},-s_t) H(s_t).$

We now have the freedom to design $H(s_t)$ in such a way that $G_f(\Vecs_{\Vecx},s_t)$ has some desirable properties. For our framework it is favourable to choose $H(s_t)$ such that $G_f(\Vecs_{\Vecx},s_t)$ decays quickly, at least in the $s_t$-domain. This will reduce the approximation error \eqref{eq:ConstantLeastSquaresCoeffs_Error} as discussed in \Cref{ssec:C6_ExactApproxStrangFix}.

\begin{figure}[tb]
  \centering
	\resizebox{\linewidth}{!}{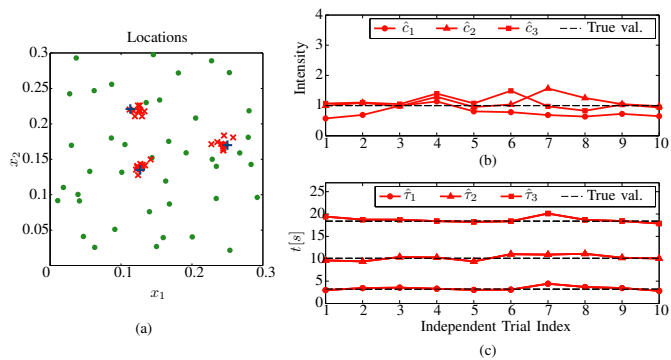}\vspace{-3mm}
\caption{\textbf{Centralised estimation of diffusion sources from noisy measurements obtained by randomly placed sensors.} The field is sampled at $1Hz$ for $T=25s$ using $41$ nonuniformly placed sensors, samples have SNR $= 15$dB. (a) Estimated locations, (b) Estimated intensities and (c) Estimated activation times, using $r=1$ and $K_1 = K_2 = 15$ for estimation algorithm.}
\label{fig:3SourceDiffSourceReconUnif_USF_7K1K2_15dB}
\end{figure}

\begin{figure}[tb]
  \centering
	\resizebox{0.8\linewidth}{!}{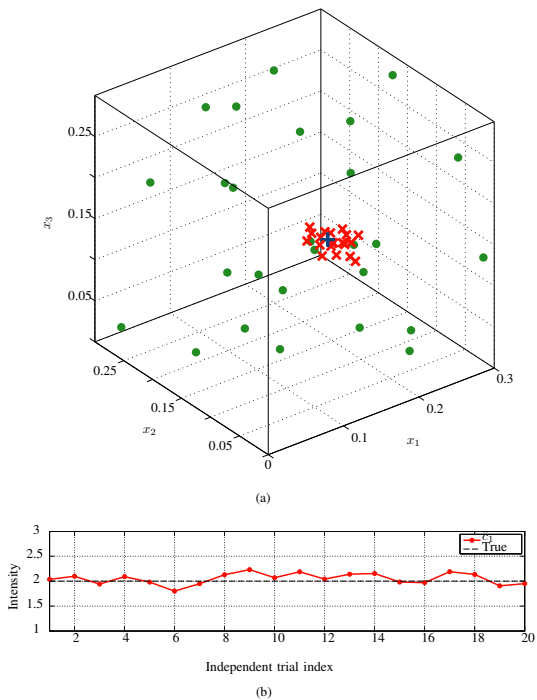}\vspace{-3mm}
\caption{\textbf{Centralised estimation of single localised source for Poisson's equation using nonuniformly placed sensors.} The field is sampled at $1Hz$ for $T=1s$ using $27$ randomly placed sensors, samples have SNR $= 15$dB. (a) Location estimates (b) Estimated intensities using $r=0$ and $K_1 = K_2 = K_3 = 2$ for the estimation algorithm.}
\label{fig:Poisson3D_Loc_Intns_27Sensors_5K1K2_20dB}
\end{figure}

\section{Numerical Simulations}
\label{sec:NumericalResults}
\begin{figure}[tb]
	\centering
	\includegraphics[width=0.65\linewidth]{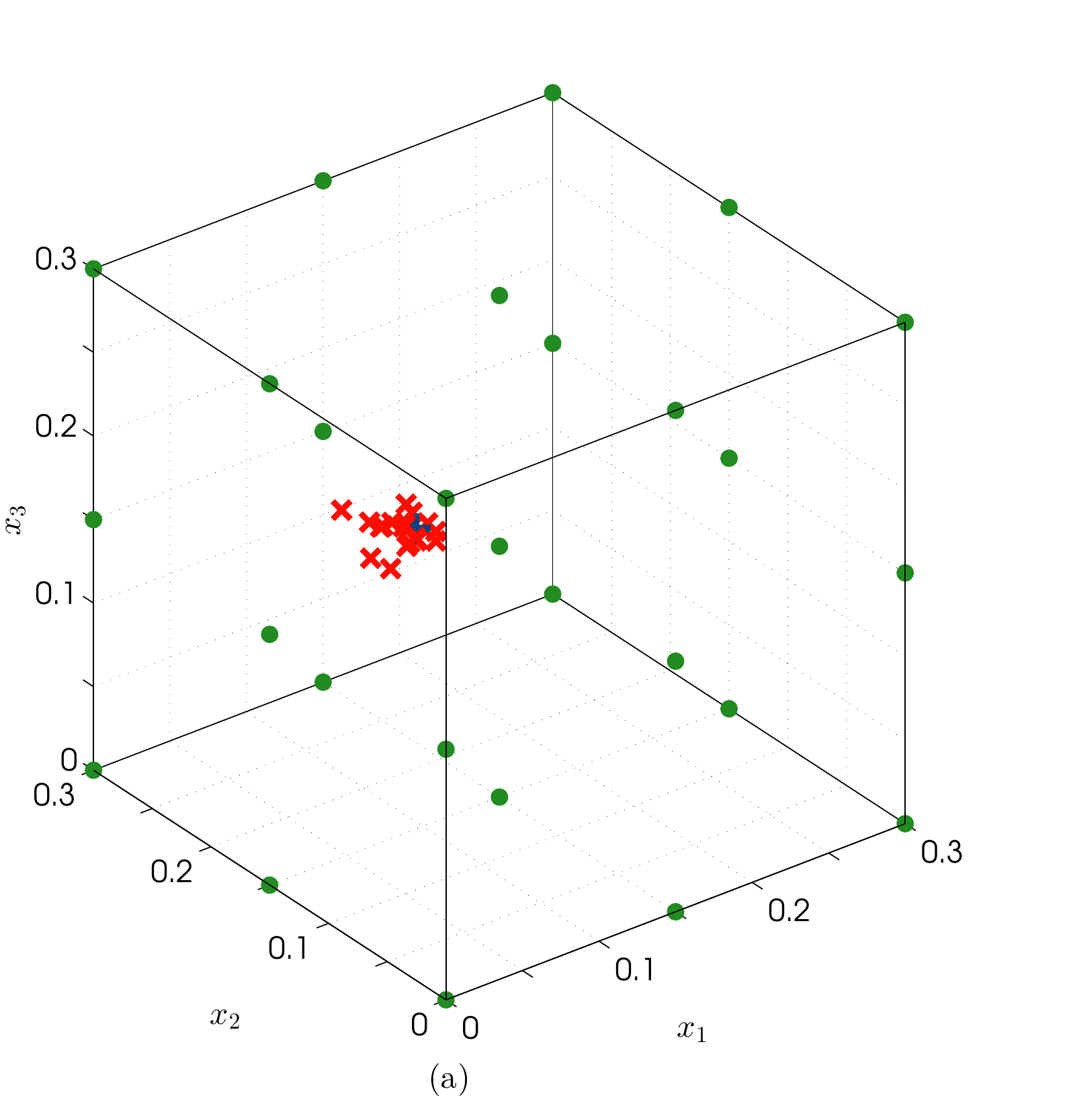}
	\includegraphics[width=0.8\linewidth]{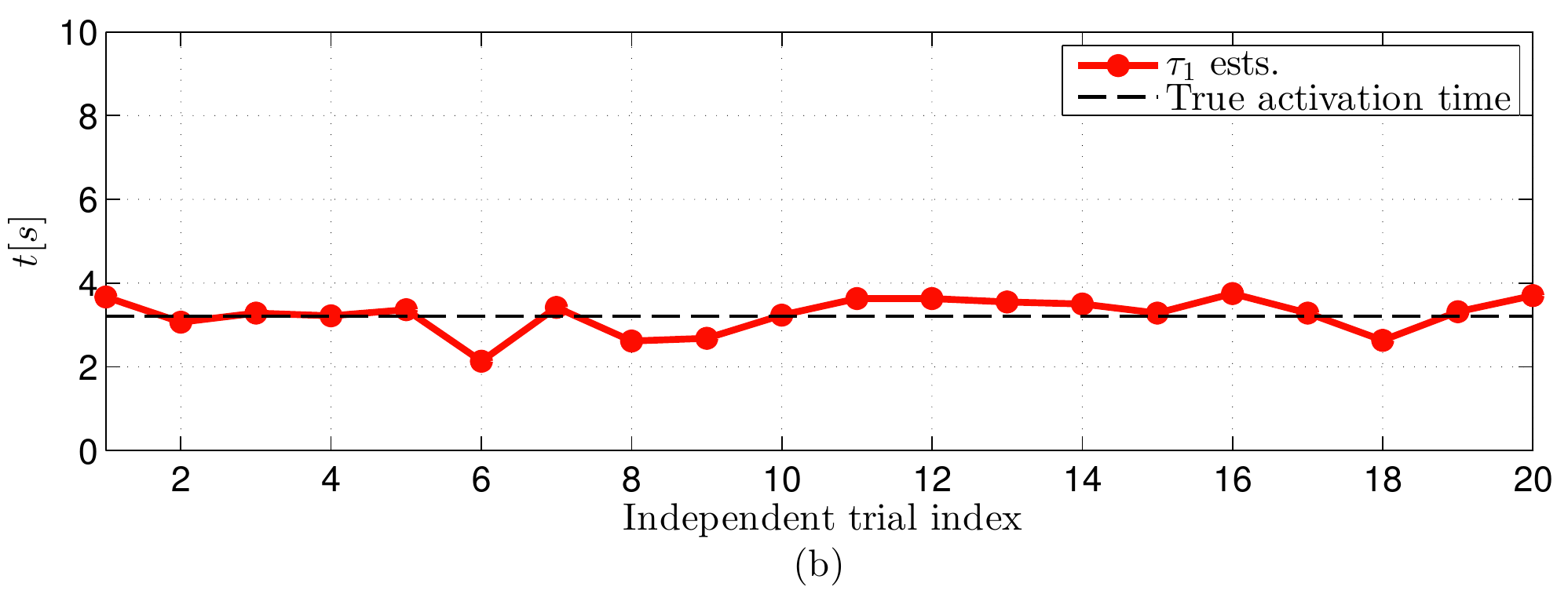}		
	\includegraphics[width=0.9\linewidth]{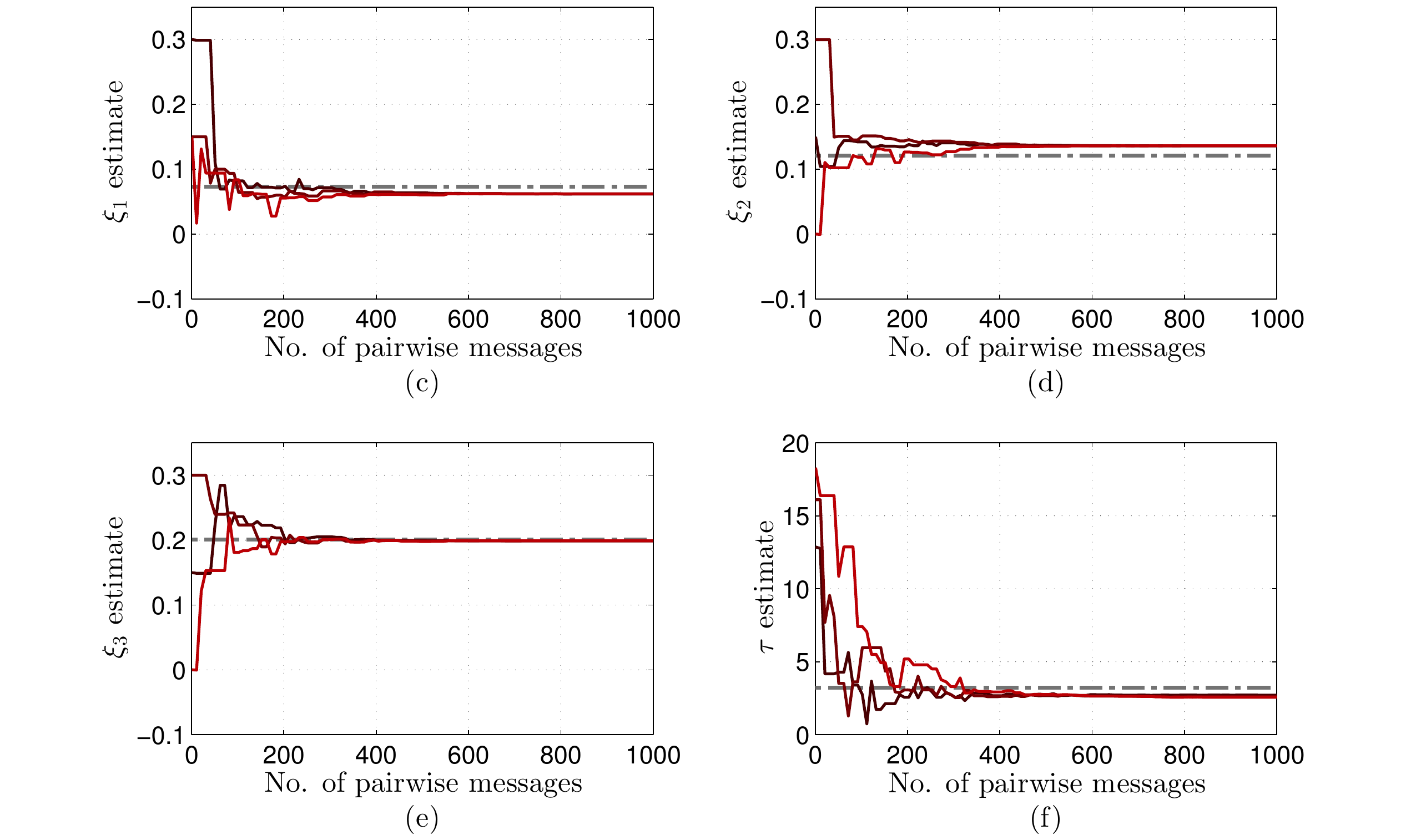}	
\caption{\textbf{Distributed estimation of localised sources for the Wave equation using uniformly placed sensors.} The field is sampled at $1Hz$ for $T=20s$ using $27$ uniformly placed sensors, samples have SNR $= 10$dB. (a) Location estimates, (b) Estimated activation times. By considering one experiment, the evolution of the estimates of (c) $\xi_1$, (d) $\xi_2$, (e) $\xi_3$, and (f) $\tau$, for three randomly chosen sensors are shown. Here we used $r=1$ and $K_1 = K_2 = K_3 = 5$ for the estimation algorithm.}
\label{fig:Wave3D1Hz20s_Loc_Actvs_27Sensors_9K1K2_15dB}
\end{figure}

In this section, we present some numerical results to validate the proposed framework for solving ISPs driven by linear PDEs. To investigate several scenarios, we present results for cases where the measured phenomena have been generated by:
\begin{enumerate}
	\item \textbf{the diffusion equation} -- we simulate a $2$-D multiple source field assuming a nonuniform sensor placement and investigate the proposed interpolation approach outlined in \Cref{sssec:Interpolation}. The results are shown in \Cref{fig:3SourceDiffSourceReconUnif_USF_7K1K2_15dB}. We also devote \Cref{ssec:EstimationPerformanceComp} to comparing the performance and computational complexity of the proposed framework to a sparse synthesis recovery scheme \cite{Kitic2016_Physics}.
	\item \textbf{Poisson's equation} -- we consider a single source field in $3$-D with nonuniform spatial samples, and utilise the linear system approach to find the desired analysis coefficients. Corresponding results are presented in \Cref{fig:Poisson3D_Loc_Intns_27Sensors_5K1K2_20dB}.
	\item \textbf{the wave equation} -- we simulate a single source wave field in $3$-D, sampled using a distributed SN of uniformly placed sensors. We assume that the sensors filter the field in time using a cubic spline before sampling. The results obtained are summarised in \Cref{fig:Wave3D1Hz20s_Loc_Actvs_27Sensors_9K1K2_15dB}.
\end{enumerate}

The sensor measurements are simulated numerically and artificially corrupted by AWGN as defined in \eqref{eq:SNR_model}, using Matlab. We then apply our estimation algorithms on the measurements. For statistical significance, multiple independent trials are performed within each setup, with both a new noise process and random sensor placement (for the nonuniform sampling setups). In \Cref{fig:3SourceDiffSourceReconUnif_USF_7K1K2_15dB,fig:Poisson3D_Loc_Intns_27Sensors_5K1K2_20dB,fig:Wave3D1Hz20s_Loc_Actvs_27Sensors_9K1K2_15dB}, the green `$\bullet$', red `$\times$' and blue `$+$' denote the sensor locations, the estimated and true source locations, respectively.

\subsection{Inverse source problem for the diffusion equation}
We present numerical results in the nonuniform sampling scenario, where the diffusion field is induced by three localised and instantaneous sources. We use the $2$-D test function family $\{\Psi_{\VecIndexk}(\Vecx) \Gamma_r(t) = e^{\imagj k_1 x_1 + \imagj k_2 x_2 } e^{\imagj r t/T}\}_{\VecIndexk,r}$ with $r = 1, k_1 =k_2 = 1, 2,\ldots, 15$ and present in \Cref{fig:3SourceDiffSourceReconUnif_USF_7K1K2_15dB} the source estimation results obtained by using \Cref{alg:UniversalSimultaneousPointSourceEst_Central} with a linear interpolator and a resampling grid with $\Delta_{x_1} = \Delta_{x_2} = 1/30$. For statistical significance, we perform 20 independent trials of the experiment.

Moreover for the $2$-D field, with Green's function \eqref{eq:GreensFunction_Diff}, we obtain its Laplace transform as (see Appendix \ref{app:C6_GaussianLT}):\vspace{-0.5mm}
\begin{equation}
		G(\Vecs_{\Vecx},s_t) = \frac{1}{s_t - \mu \|\Vecs_{\Vecx}\|^2},\vspace{-0.5mm}
\label{eq:LaplaceTransform_Diffusion}
\end{equation}
provided $\Re{\left( s_t - {\mu}\|\Vecs_{\Vecx}\|^2 \right)} > 0 $, where $\Re{\left( z \right)} $ is used to denote the real part of a complex number $z$. Hence, by substituting \eqref{eq:LaplaceTransform_Diffusion} into \eqref{eq:ExpReproducingCoeffs_NonInteger} with $\bm{\kappa} = \imagj(\VecIndexk, r/T)$, the desired exponential reproducing coefficients are of the form:\vspace{-0.5mm}
\begin{equation}
w_{\VecIndexn,l}(\VecIndexk,r) = \frac{\Delta_{x_1} \Delta_{x_2} \Delta_{t}(\mu(k_1^2 + k_2^2 ) + \imagj r/T )}{e^{-\imagj (\Delta_{x_1}k_1 n_1 + \Delta_{x_2}k_2 n_2 + \Delta_t r l/T)}}.\vspace{-0.5mm}
\label{eq:DiffCoeffs}
\end{equation}

As seen in \Cref{fig:3SourceDiffSourceReconUnif_USF_7K1K2_15dB}, all source parameters have been recovered reliably; in particular, the source locations (\Cref{fig:3SourceDiffSourceReconUnif_USF_7K1K2_15dB}(a)) and activation times (\Cref{fig:3SourceDiffSourceReconUnif_USF_7K1K2_15dB}(b)), which are usually the two main parameters of interest, vary only marginally around their true values despite the low measurement SNR.
\begin{rem}
The requirement that $\Re{\left( s_t - {\mu}\|\Vecs_{\Vecx}\|^2 \right)} > 0 $ is necessary for the transform integral to converge. By substituting $(\Vecs_{\Vecx}, s_t) = -\imagj(\VecIndexk,r/T)$ into \eqref{eq:LaplaceTransform_Diffusion}, it is easy to see that this is satisfied when $ \VecIndexk \neq \{\bm{0}\}$.
\end{rem}

\subsection{Inverse source problem for Poisson's Equation}
Since the Poisson field is static, we only focus on recovering the source location and intensity. Specifically, we estimate the unknown source from a single time-snapshot of the $3$-D potential field measurements by formulating the linear system discussed in \Cref{ssec:ExplicitCoefficients}, with $r=0$ and $K_1 = K_2 = K_3 = 2$. Therefore according to \eqref{eq:DirectMethod_LinearSys_Space}, we get:
\begin{displaymath}\vspace{-2mm}
\left[ \begin{array}{cccc}
	g(\Vecx_1 {-} \Vecx'_1)  & \cdots & g(\Vecx_N {-} \Vecx'_1) \\
	g(\Vecx_1 {-} \Vecx'_2)  & \cdots & g(\Vecx_N {-} \Vecx'_2) \\
					 \vdots							      & \ddots & 					\vdots								 \\
	g(\Vecx_1 {-} \Vecx'_I)  & \cdots & g(\Vecx_N {-} \Vecx'_I)
\end{array} \right] \! \!
\left[\arraycolsep=0.015pt\def\arraystretch{1.2} \begin{array}{c}
	w_{1,l}(\VecIndexk,0) \\
	w_{2,l}(\VecIndexk,0) \\
	\vdots \\
	w_{N,l}(\VecIndexk,0)
\end{array} \right] \!
 {=} \! \left[\arraycolsep=0.015pt\def\arraystretch{1.2} \begin{array}{c}
	\Psi_{\VecIndexk} (\Vecx'_1) \\
	\Psi_{\VecIndexk} (\Vecx'_2) \\
	\vdots \\
	\Psi_{\VecIndexk} (\Vecx'_I)
\end{array} \right],
\end{displaymath}
where $I = 1000$ and $\Vecx'_{i}$ is obtained from a lexicographic ordering of $\{ (i_1 \delta_{x_1} + \varepsilon, i_2 \delta_{x_2} + \varepsilon, i_3 \delta_{x_3} + \varepsilon ) \}_{i_1,i_2,i_3 = 0}^9$ with $\delta_{x_1} = \delta_{x_1} = \delta_{x_1} = 0.03$,\footnote{The slight shift $\varepsilon \geq 0$ is used here to avoid the singularity of $g(\Vecx)$ at $\|\Vecx\| = 0$.} and $g(\Vecx) = - \frac{1}{4 \pi \|\Vecx\|}$.

The simulation results are shown in \Cref{fig:Poisson3D_Loc_Intns_27Sensors_5K1K2_20dB}, with the source location and intensity estimates obtained over the $20$ independent trials. All estimates show only small variations around the ground truth parameters. In particular, the variation in location estimates is much smaller than the average sensor spacing. Furthermore the condition number for the matrix $\VecG$ is reasonably small falling in the range $15$ -- $ 38$.

\begin{figure}[htb]%
\centering
\includegraphics[width=0.7\linewidth]{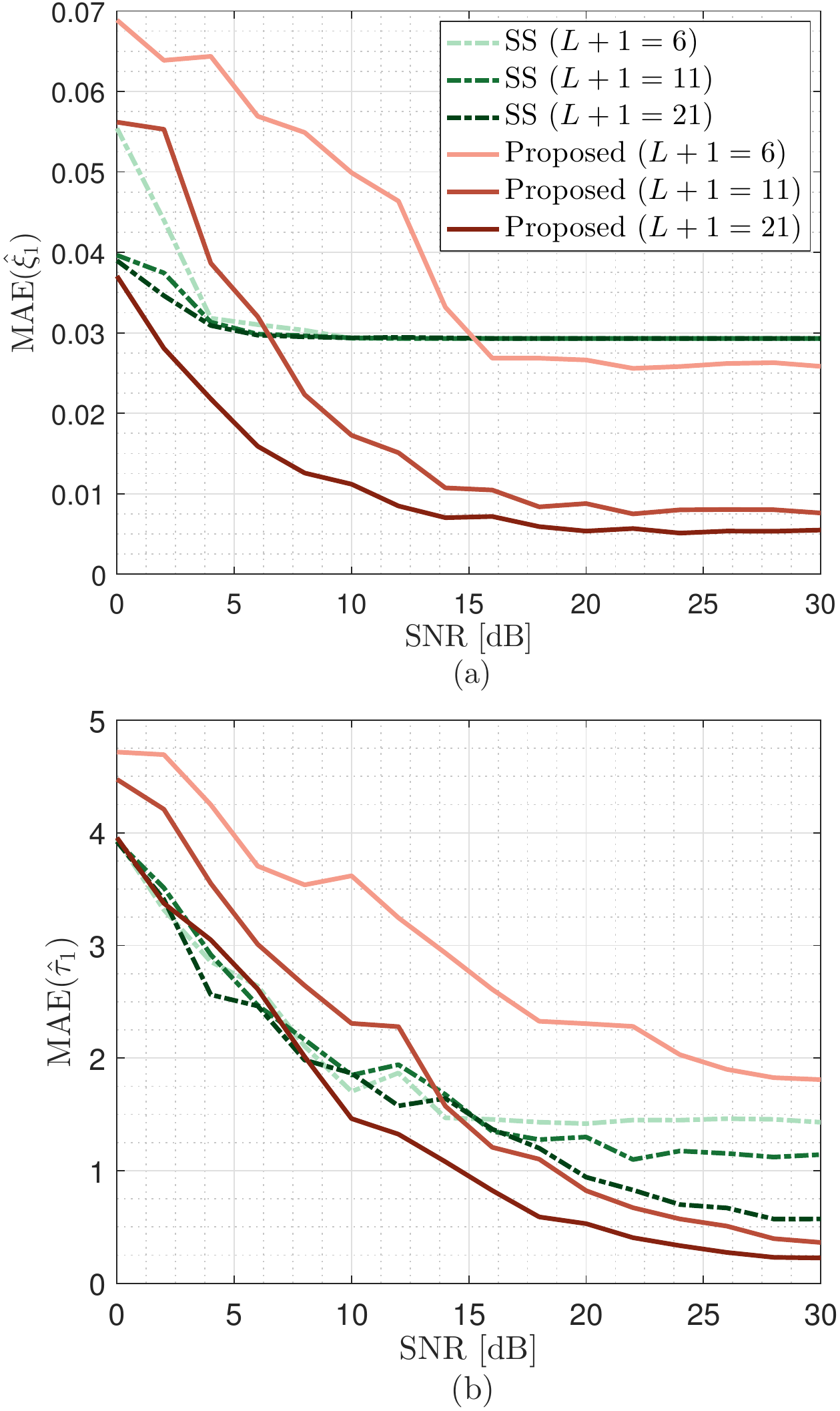}%
\caption{\textbf{Noise robustness.} Mean absolute error of the estimated (a) locations and (b) activation times against SNR, for $N=6$ sensors and $L+1 = \{6, 11, 12\}$ temporal samples.}%
\label{fig:EstPerformanceSNR_5Sensors}%
\end{figure}%
%
%
%
%
\begin{figure*}[htb]%
\centering
\includegraphics[width=0.75\linewidth]{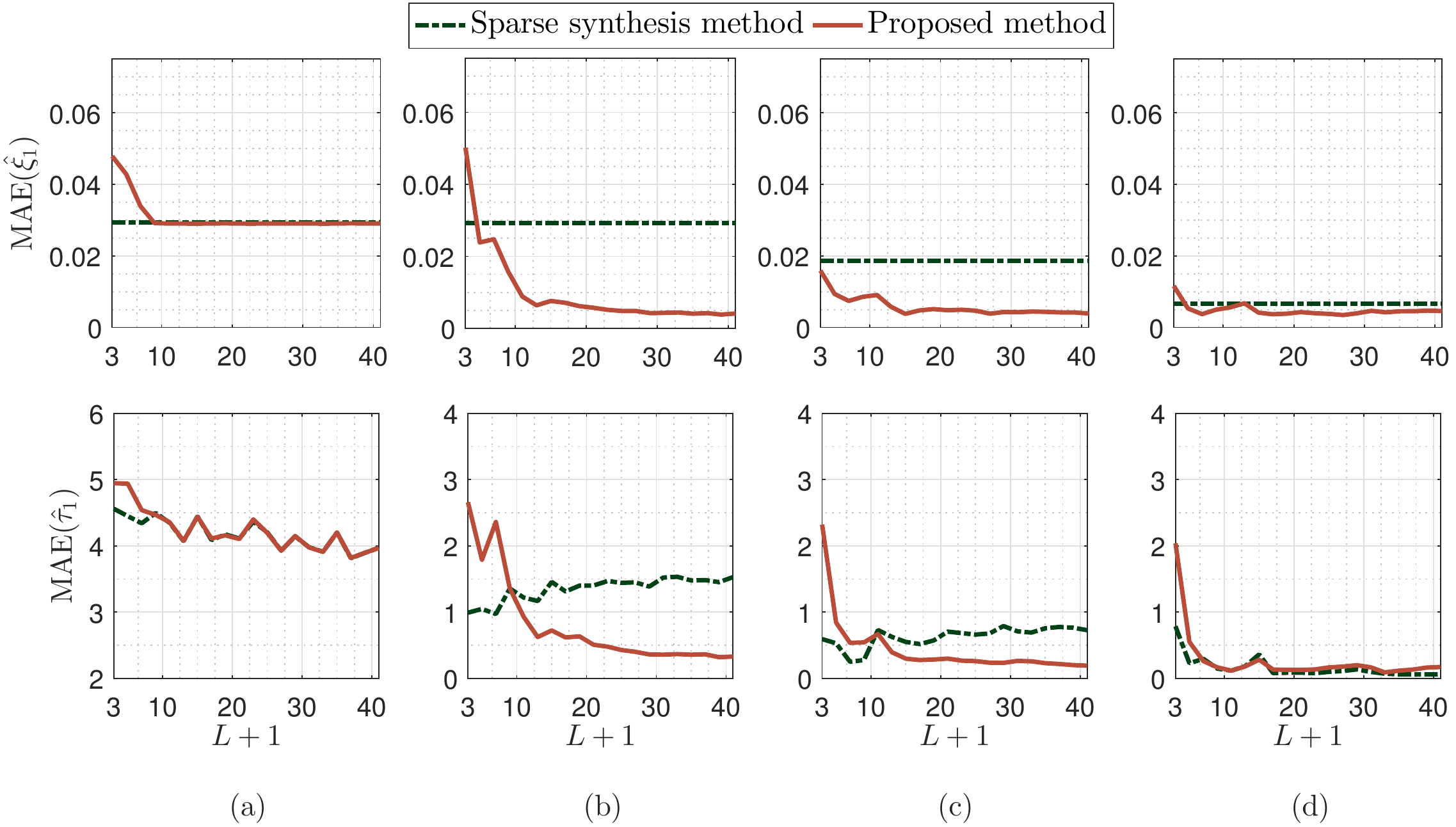}%
\caption{\textbf{Estimation performance.} Mean absolute error of the estimated locations (top) and activation times (bottom) against the number of temporal samples $L+1$ for different number of sensors: (a) $N=3$, (b) $N=5$, (c) $N=7$, (d) $N=9$; measurement SNR$=20$dB.}%
\label{fig:EstPerformance_NumMeas20dB}%
\end{figure*}%
\begin{figure*}[htb]
\centering
\includegraphics[width=0.75\linewidth]{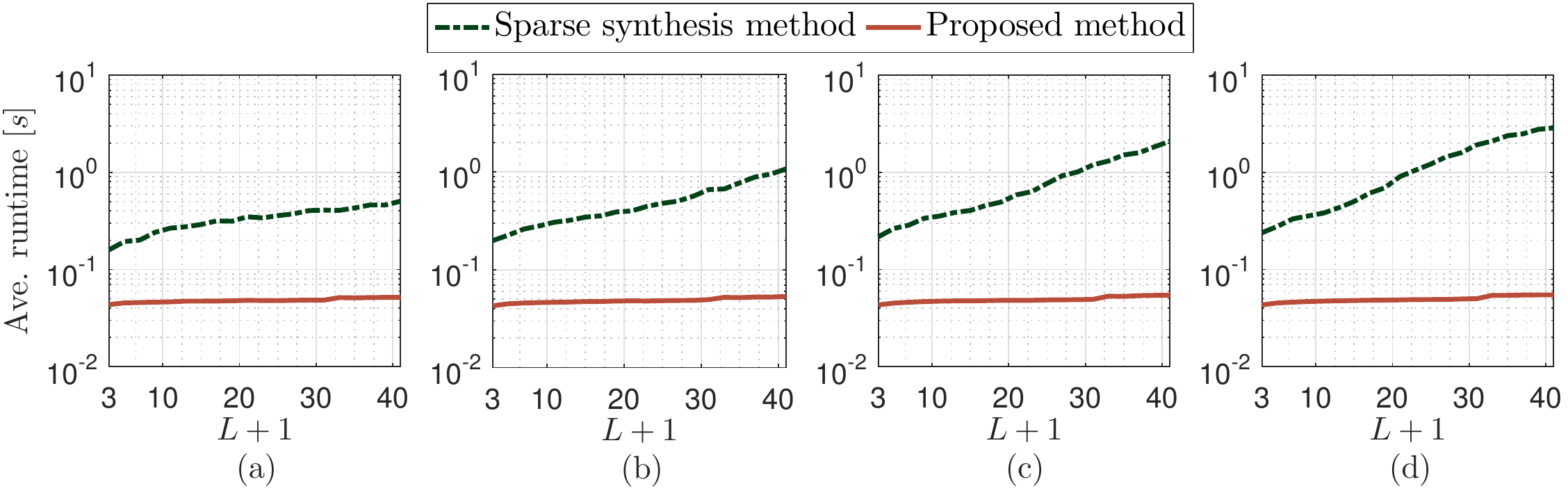}%
\caption{\textbf{Computation time.} Average runtime for each method against $L+1$ for different number of sensors: (a) $N=3$, (b) $N=5$, (c) $N=7$, (d) $N=9$; measurement SNR$=20$dB.}%
\label{fig:RuntimePerf_20dB}%
\end{figure*}%
%

\subsection{Distributed acoustic source localisation: inverse source problem for the wave equation}
Here the simulated physical phenomena is a $3$-D wave field, obeying \eqref{eq:WaveEqn} and, induced by a single source. We consider the distributed estimation setup, wherein the SN model is assumed to be the RGG $\mathcal{G}(N,0.4)$. Each of the $N=27$ regularly placed sensors acquires the field by filtering it (in time) using a third order B-spline before sampling. We apply the approach described in \Cref{ssec:TemporalFiltering} by first computing the desired coefficients $w_{\VecIndexn,l}(\VecIndexk,r)$ from the filtered Green's function $g_f(\Vecx,t)$, which has the bilateral Laplace transform, $G_f(\Vecs_{\Vecx}, s_t) = G(\Vecs_{\Vecx}, s_t)H( s_t)$, where $H(s_t) = \left(\frac{1 - e^{-s_t}}{s_t}\right)^4,$ is the Laplace transform for the third order B-spline\footnote{The zero order B-spline is taken as the indicator function on [0,1], from this we define the $n$-th order B-spline as the spline generated by convolving $(n+1)$ zero order B-splines.}, whereas $G(\Vecs_{\Vecx}, s_t) = \frac{1}{ \|\Vecs_{\Vecx}\|^2 - ({s_t/c})^2 }$. A derivation of this expression is provided in Appendix \ref{app:C6_WaveLT}. Combining these gives us the desired coefficients,
\begin{equation}
w_{\VecIndexn,l}(\VecIndexk,r) = \frac{\Delta_{t}\prod_{i=1}^3\Delta_{x_i} (r/T)^4 \left( (r/cT)^2 - \lVert \VecIndexk \rVert^2 \right)}{(1 - e^{-\imagj r/T })^4 e^{-\imagj( \sum_{j=1}^3 k_j n_j \Delta_{x_j} + r l \Delta_{t}/T)} },
\label{eq:WaveFieldCoeffs}
\end{equation}
where $ \Delta_{x_1} =  \Delta_{x_2} =  \Delta_{x_3} = 0.1$ and $ \Delta_{t} = 1$. Moreover,  we choose $K_1 =  K_2 =  K_3 = 5$ and $r=1$ for the sensing function family $\{ \Psi_{\VecIndexk}(\Vecx) \Gamma_r (t) = e^{k_1 x_1 + \imagj k_1 x_2 + \imagj k_2 x_3 } e^{\imagj r t/T}\}_{\VecIndexk,r}$. Given these coefficients, the sensor nodes can perform a distributed estimation of the unknowns via gossiping, as described in \Cref{ssec:Distr_SNalgorithm}. The estimation results for 20 independent trials in this uniform sampling case is presented in \Cref{fig:Wave3D1Hz20s_Loc_Actvs_27Sensors_9K1K2_15dB}, wherein the recovered source locations, in (a), and activation times, in (b), are plotted over the true values. From these plots, we can conclude that the estimates are reliable, even when we have noisy measurements. In addition by considering a single experiment only, \Cref{fig:Wave3D1Hz20s_Loc_Actvs_27Sensors_9K1K2_15dB}(c)--(f) shows the evolution of the location and activation time estimates for three randomly chosen sensors (with each pairwise gossip round). Specifically, \Cref{fig:Wave3D1Hz20s_Loc_Actvs_27Sensors_9K1K2_15dB}(c), (d) and (e) displays the location estimates (in each spatial dimension), i.e. $x_1, x_2$ and $x_3$ respectively. Similarly, \Cref{fig:Wave3D1Hz20s_Loc_Actvs_27Sensors_9K1K2_15dB}(f) shows the evolution of the activation time estimate. In both cases we notice that the estimates converge to the true value, despite the low measurement SNR.
\subsection{Estimation performance comparison}
\label{ssec:EstimationPerformanceComp}
We now numerically compare the proposed scheme against the sparse synthesis (SS) formulation described in \cite{Kitic2016_Physics}. For the SS method, we formulate the LASSO problem:\vspace{-2mm}
\begin{equation}
\min_{\Vecf} \frac12 \|\VecD\Vecf - \bm{\varphi}\|_2 + \gamma \|\Vecf\|_1,\vspace{-2mm}
\label{eq:LassoProblem}
\end{equation}
where the dictionary $\VecD \in \mathbb{R}^{N(L+1) \times N_x N_t}$ is formed by discretising the Green's function appropriately assuming uniform spatial and temporal grids with $N_x = 51$ and $N_t=101$ divisions, $\Vecf\in \mathbb{R}^{N(L+1)}$ is the source discretisation, whilst $\bm{\varphi}\in \mathbb{R}^{N_xN_t}$ is the vectorised sensor samples. The problem is then solved using the alternating direction method of multipliers (ADMM) \cite{Boyd2011_Distributed}. The sparse synthesis formulation exhibits the same estimation performance as the sparse analysis formulation, proposed in \cite{Albera2014_Brain,Kitic2016_Physics}, so we only focus on one of them; although SS is much more computation intensive. The proposed sampling-based method uses $r=1$ and $K=50$. Note that for both approaches their activation time estimates are further refined by performing a line search in a small neighbourhood of the initial estimate.

The field used is $1$-D diffusion induced by a single source with parameters $c_1 = 5$, $\xi_{1}=0.1207m$ and $\tau_1 = 1.2175s$ over $x \in \Omega = [0,0.3]m$ and $t \in [0,20]s$. All reported statistics were computed from $200$ independent trials, each having a new noise realisation.

\Cref{fig:EstPerformanceSNR_5Sensors} shows that for very low SNR combined with low samples $N=6$ and $L+1=6$, the sparsity-based recovery method achieves a lower mean absolute error (MAE) than the proposed method. However at higher temporal sampling frequency, for e.g. $L+1=21$, the current method consistently outperforms the SS method even at an extremely low SNR$=0$dB. Moreover, for fixed SNR $=20$dB we observe, in \Cref{fig:EstPerformance_NumMeas20dB}, that the MAE remains constant even as the number of temporal samples increases -- largely due to the discretisation; where as the current approach improves drastically with the number of temporal samples, outperforming the SS method for most values of $N$ and $L$. Finally by comparing the runtimes of both schemes in \Cref{fig:RuntimePerf_20dB}, we notice that the complexity of the present method increases only marginally, whilst the SS method measures around one to two orders of magnitude slower as the number of temporal and spatial samples increase.
%
%
%
%
\section{Conclusion}
\label{sec:Conclusion}
In this work we established a new general framework to solve ISPs for fields described by linear PDEs with constant coefficients. This formulation reduced the ISPs, for such physics-driven fields, to the problem of reproducing exponentials using shifted versions of the space- and time-reversed Green's function of the corresponding field, under the assumption that the sources are localised. Consequently, we proposed practical sensor network algorithms for both uniform and nonuniform sampling setup. Finally, we validated our framework on some popular PDE models encountered in various applications, and also demonstrated that our method compares favourably against sparse recovery based methods.

%
%
\appendices
\section{Estimating Parameters of Multidimensional Superimposed Exponentials}
\label{app:ParamEstimation_SoS}
Consider the sequence (of generalised measurements) 
\begin{equation}
	\mathcal{Q}(\VecIndexk,r) = \mathcal{Q}(k_1,\ldots,k_d,r) = \sum_{m=1}^M c_m b_m^r \prod_{i=1}^d(v_{i,m})^{k_i},
	\label{eq:Appendix_PowerSumSeries}
\end{equation}
for any $r$, it is a superposition of $M$ $d$-dimensional damped complex sinusoids. Moreover, $k_i = 0, 1, \ldots, K_i - 1$ for each $i =1,2,\ldots ,d$. This multidimensional Prony-like system is prevalent in spectral estimation and array processing applications for example. Many approaches have been put forward for recovering the unknown frequencies and amplitudes, particularly for $d=2$ \cite{Hua1990_Matrix,Rouquette2001_Estimation,Maravic2004_Exact}.

In our framework, discussed in \Cref{sec:SensingFunctions}, solving this system for $\{b_m, c_m, (v_{i,m})_{i=1}^d\}_{m=1}^M$ allows us to recover the unknown source parameters $\{c_m, \tau_m,\bm{\xi}_m\}_{m=1}^M$. We utilise the $N$-D ESPRIT algorithm by Sahnoun \textit{et al} \cite{Sahnoun2016_Multidimensional} (based on the $2$-D ESPRIT algorithm of \cite{Rouquette2001_Estimation}) to solve this system. The $N$-D ESPRIT algorithm is as follows:
\begin{enumerate}
	\item Choose $L_1,L_2, \ldots, L_d \in \mathbb{N}$ such that $1 \leq L_i \leq K_i$ and set $J_i = K_i - L_i + 1$.
	\item Construct the multilevel Hankel matrix $\VecH$
	\begin{equation}
		\begin{aligned}
		\VecH = [&\Vecq_{1, \ldots, 1} \, \Vecq_{1, \ldots, 1, 2} \, \cdots \, \Vecq_{1, \ldots, 1, J_d} \\
									&\Vecq_{1, \ldots, 1,2,1} \, \cdots \\
									& \cdots \, \Vecq_{J_1, J_2, \ldots, J_{d-1}} \, \Vecq_{J_1, J_2, \ldots, J_d} ],
		\end{aligned}
		\label{eq:MultiLevelHankel}		
	\end{equation}
	where the vectors $\Vecq_{\VecIndexj} {\defeq} \mathrm{vec}\mathopen{}\left( \mathcal{Q}(\VecIndexj-\bm{1}:\VecIndexj+\VecIndexL-2\cdot\bm{1},r) \right)$, $\VecIndexj {=} (j_1, j_2, \ldots, j_d) \in \mathbb{N}^d_{+}$, and $\VecIndexL = (L_1, L_2, \ldots, L_d)$. Here $\VecIndexj:\VecIndexj+\VecIndexL-\bm{1} \defeq (j_1:j_1 +L_1 - 1, j_2:j_2 +L_2 - 1, \ldots, j_d:j_d +L_d - 1)$ is the (MATLAB-like) notation used for extracting subarrays, and $\mathrm{vec}\mathopen{}( \cdot \mathclose)$ is the vectorisation operator.
	\item Retrieve the singular value decomposition (SVD) of $\VecH$, and form $\VecU \in \mathbb{C}^{(L_1 L_2 \cdots L_d)\times M}$, which is a matrix of the $M$ most dominant left singular vectors of $\VecH$.
	\item Compute the matrices $\VecF_i$, using\vspace{-1mm}
	\begin{equation}
		\VecF_{i} =\left( _i\underline{\VecU} \right)^\dagger \left(^i\overline{\VecU}\right),\vspace{-1mm}
	\label{eq:Fmat}
	\end{equation}
	where $_i\underline{\VecU} = \left. _i\underline{\VecI}\right.\VecU$ and $ ^i\overline{\VecU} = \left.^i\overline{\VecI}\right.\VecU$. Moreover, $_i\underline{\VecI} = \VecI_{\prod_{j=1}^{i-1} L_j} \otimes \underline{\VecI}_{L_i} \otimes \VecI_{\prod_{j=i+1}^{d} L_j}$ and $^i\overline{\VecI} = \VecI_{\prod_{j=1}^{i-1} L_j} \otimes \overline{\VecI}_{L_i} \otimes \VecI_{\prod_{j=i+1}^{d} L_j}$. The symbol $\otimes$ denotes the Kronecker product, $\VecI_n$ is the $n \times n$ identity matrix and the overbar (respectively underbar) represents the operation of deleting the first (respectively last) row of a matrix.
	\item For some random choice of $\beta_1, \beta_2, \ldots, \beta_d$, compute the linear combination of matrices:\vspace{-1mm}
	\begin{equation}
		\VecK = \sum_{i=1}^d \beta_i \VecF_i.\vspace{-1.5mm}
	\label{eq:LinCom}
	\end{equation}
	\item Diagonalise the matrix $\VecK$ to find $\VecT$, such that\vspace{-1.5mm}
	\begin{equation}
		\VecK = \VecT \mathrm{diag}\mathopen{}\left( \bm{\eta} \right) \VecT^{-1}.\vspace{-1.5mm}
	\label{eq:Diag1}
	\end{equation}
	\item Then for each $i = 1, 2, \ldots, d$, transform $\VecF_i$ using the matrix $\VecT$,\vspace{-1.5mm}
	\begin{equation}
			\VecD_i = \VecT^{-1} \VecF_i \VecT.\vspace{-1.5mm}
	\end{equation}
	\item Compute the unknown frequencies $\{ v_{i,1}, v_{i,2}, \ldots, v_{i,M} \}$ as the diagonal of $\VecD_i$, i.e. $\mathrm{diag}\mathopen{}\left( \VecD_i \right)$ for each $i=1,2,\ldots,d$.
	\item Using $\{ v_{i,1}, v_{i,2}, \ldots, v_{i,M} \}_{i=1}^d$, solve the least-squares problem associated with \eqref{eq:Appendix_PowerSumSeries} to find $\{c_m' = c_mb_m^r\}_{m=1}^M$.
\end{enumerate}

\section{Gossiping for Distributed Average Consensus}
\label{app:GossipAlgorithms}
The problem of achieving consensus or agreement amongst agents of a network in a distributed fashion, is well-studied. Gossip algorithms, based on the early works of Tsitsiklis \textit{et al} \cite{tsitsiklis1984problems}, have been applied to the distributed averaging problem, as they possess the attractive property of not requiring a specialised routing strategy. Consider for example pairwise randomised gossip \cite{Boyd2006_Randomized}, which involves at each time step two random but connected nodes updating their values with a weighted average of their current values.

Let us denote the value at node $n$ after the $i$-th pairwise gossip by $y_{n,i}$, then the initial value is $y_{n,0}$. At each iteration, a random node $n$ wakes up and contacts a randomly chosen neighbor $n'$, they both update their estimates with $y_{n,i{+}1} = y_{n',i{+}1} = (y_{n,i} {+} y_{n',i})/2$. Furthermore, let $\Vecy(i) = [y_{1,i}, y_{2,i}, \ldots, y_{N,i}]\transpose$ then this pairwise gossip algorithm can be summarised mathematically as:
\begin{equation}
\Vecy (i+1) = \VecP(i) \Vecy(i)
\label{eq:PairwiseGossip}
\end{equation}
where $\VecP(i)$'s are doubly stochastic matrices drawn randomly at the $i$-th iteration. Moreover, $\VecP(i)$ is a diagonal identity matrix everywhere else apart from the elements $(n,n),(n,n'),(n',n)$ and $(n',n')$ which are all equal to $1/2$.

If the network is connected and the nodes communicate using this scheme, then it can be shown that each node is guaranteed to converge to the global network average $\bar{y} = \frac{1}{N} \sum_{n=1}^N y_{n,0}$, after enough iterations, i.e. $\lim_{i\rightarrow \infty} \Vecy(i) = \mathbf{1}\bar{y}$. Performance guarantees and convergence results have also been studied (see \cite{Boyd2006_Randomized} and the references therein).

The \textit{localised interactions} in our inverse source problem set up will be based on the use of gossip algorithms to compute the generalised measurements $\{\mathcal{Q}(\VecIndexk,r)\}_{\VecIndexk} = \sum_{n,l} w_{n,l}(\VecIndexk,r) \varphi_{n}(t_l)$, in a distributed manner. Specifically, we assume each node $n$ knows only its sequence of coefficients $\{w_{n,l}(k,r)\}_{l}$ and senses the field $\varphi_{n}(t_l)$ in time, then the sensors communicate their local measures $y_n(\VecIndexk,r) = N \sum_l w_{n,l}(\VecIndexk,r) \varphi_n(t_l)$. On convergence, the nodes will have $\mathcal{Q}(\VecIndexk,r) = \frac{1}{N} \sum_{n=1}^N y_n(\VecIndexk,r),$ which is the desired generalised measurements.

\section{The Generalised Strang-Fix Conditions}
\label{app:C6_GenStrangFix}
We first state the classical Strang-Fix condition \cite{Strang1973_Fourier}, for multidimensional polynomial reproduction.
\begin{myLem}[Strang-Fix condition \cite{Strang1973_Fourier,Uriguen2013FRI}]
Any compactly supported kernel $\psi(\Vecx)$ whose derivatives up to and including order $p$ are in $\Elltwo$, is able to reproduce polynomials, i.e.:
$$\Vecx^{\bm{\alpha}} = \sum_{\VecIndexn \in \mathbb{Z}^d} c_{\bm{\alpha},\VecIndexn} \psi(\Vecx-\VecIndexn),$$
where $\bm{\alpha} = (\alpha_1, \ldots, \alpha_d)$ with $\sum_i \alpha_i \leq p$, if and only if
$$\hat\psi(\bm{0}) \neq 0 \; \text{ and } \; \nabla^{\bm{\alpha}} \hat\psi(2\pi \bm{\ell}) = 0,$$
where $\hat\psi(\imagj \bm{\omega})$ is the Fourier transform of $\psi$, and $\bm{\ell} \in \mathbb{Z}^d\setminus\{\bm{0}\}$.
\begin{IEEEproof}
The proof of this lemma depends on the Poisson summation formula -- which connects the values of a function $\psi$ on the lattice $\mathbb{Z}^d$ with its Fourier transform $\hat\psi$ on the lattice $2\pi \mathbb{Z}^d$. Note that both sides of the summation converge (absolutely) when $\psi$ and $\hat\psi$ decay sufficiently quickly. The proof is based on taking the Fourier transform of $\psi(\VecIndexn) = \VecIndexn^{\bm{\alpha}} \varphi(\Vecx - \VecIndexn)$ and applying the frequency differentiation property to it. The reverse implication follows immediately from this expression, whereas the forward follows by induction. A complete proof appears in \cite{Strang1973_Fourier}.
\end{IEEEproof}
\end{myLem}
To demonstrate the multidimensional generalised Strang-Fix condition, we adapt the proof of \cite{Uriguen2013FRI}. First, we require that the function $\psi(\Vecx) = e^{-\Veckappa \cdot \Vecx} s(\Vecx)$ is able to reproduce the polynomial $\Vecx^{\bm{\alpha}}$. Consequently, we obtain
\begin{align}
\Vecx^{\bm{\alpha}} &= \sum_{\VecIndexn} c_{\bm{\alpha},\VecIndexn} e^{-\Veckappa\cdot(\Vecx - \VecIndexn)} s(\Vecx) \\
\Leftrightarrow \Vecx^{\bm{\alpha}}e^{\Veckappa\cdot\Vecx} &= \sum_{\VecIndexn} c_{\bm{\alpha},\VecIndexn} e^{\Veckappa\cdot\VecIndexn} s(\Vecx),
\label{eq:Eqn1}
\end{align}
such that $w_{\VecIndexn} = \left.c_{\bm{\alpha},\VecIndexn}e^{\Veckappa\cdot\VecIndexn}\right\rvert_{\bm{\alpha} = \bm{0}}$. Moreover, it follows that \eqref{eq:Eqn1} holds true provided $\psi(\Vecx)$ satisfies the classical Strang-Fix condition with $\bm{\alpha} = \bm{0}$. This is the case when $\hat{\psi}(\bm{0}) \neq \bm{0}$ and $\hat{\psi}(2 \pi ) = \bm{0}$. Consequently, since the Fourier transform of $\psi$ is by construction related to the Laplace transform of $s(\Vecx)$, by $\hat{\psi}(\bm{\omega}) = \left. S(\Veckappa + \Vecs_{\Vecx})\right\rvert_{\Vecs_{\Vecx} = \imagj \bm{\omega}}$, we obtain the desired conditions:
\begin{equation}
	S(\Veckappa) \neq 0 \text{, and } S(\Veckappa + \imagj 2\pi \VecIndexn ) = 0.
\label{eq:GSFConditionsApp}
\end{equation}
\section{Bilateral Laplace transforms}

\subsection{Green's function of the two-dimensional diffusion equation}
\label{app:C6_GaussianLT}
Herein, we derive expression \eqref{eq:LaplaceTransform_Diffusion}, i.e. the multidimensional bilateral Laplace transform of the diffusion Green's function, $g(\Vecx,t) = \frac{1}{(4 \pi \mu t)^{\frac{d}{2}}} e^{ - \frac{\|\Vecx\|^2}{4 \mu t}} H(t).$ Thus,
\begin{equation}
 G(\Vecs_{\Vecx},s_t) {=} \!\!\int_{\Vecx \in \mathbb{R}^d} \! \! \int_{t \in \mathbb{R}} \frac{1}{(\!4 \pi \mu t\!)^{\frac{d}{2}}} e^{ {-} \frac{\|\Vecx\|^2}{4 \mu t}} H(t) e^{ \!{-} (\Vecx,t)\cdot(\!\Vecs_{\Vecx}\!, s_t\!)} \mathrm{d}t \mathrm{d}\Vecx.
\label{eq:LT_HeatKernel}
\end{equation}
We first consider the spatial integral:
\begin{align}
\int_{\Vecx \in \mathbb{R}^d} e^{ - \frac{\|\Vecx\|^2}{4 \mu t}} e^{- \Vecx \cdot \Vecs_{\Vecx}} \mathrm{d}\Vecx &= \int_{\Vecx \in \mathbb{R}^3} e^{ - \frac{\|\Vecx\|^2 + 4 \mu t \Vecx\cdot\Vecs_{\Vecx}}{4 \mu t}} \mathrm{d}\Vecx \nonumber \\
		&= \int_{\Vecx \in \mathbb{R}^3} e^{ - \frac{\|\Vecx + 2 \mu t \Vecs_{\Vecx} \|^2 - 4 \mu^2 t^2 \| \Vecs_{\Vecx}\|^2}{4 \mu t}} \mathrm{d}\Vecx \nonumber \\  
		&= e^{ \mu t \|\Vecs_{\Vecx}\|^2} \int_{\Vecx \in \mathbb{R}^3} e^{ - \frac{\|\Vecx + 2 \mu t \Vecs_{\Vecx} \|^2}{4 \mu t}} \mathrm{d}\Vecx \nonumber \\
		&= e^{ \mu t \|\Vecs_{\Vecx}\|^2} (4 \pi \mu t)^{d/2},
		\label{eq:LT_timeHeat}
\end{align}
where the second equality follows by completing the square and \eqref{eq:LT_timeHeat} uses the fact that $\int_{x=-\infty}^\infty e^{- \frac{x^2}{a}} \mathrm{d}x = \sqrt{a \pi},$ thus,
\begin{displaymath}
	\int_{\Vecx} e^{- \frac{\|\Vecx\|^2}{a}} \mathrm{d}\Vecx = \prod_{i=1}^d \int_{x_i=-\infty}^\infty e^{- \frac{x_i^2}{a}} \mathrm{d}x_i = (\sqrt{a \pi})^d,
\end{displaymath}
with $a = 4\mu t$. We can now substitute \eqref{eq:LT_timeHeat} back into \eqref{eq:LT_HeatKernel} and proceed as follows:
\begin{align}
 G(\Vecs_{\Vecx},s_t) &{=} \int_{t \in \mathbb{R}} \! e^{ \mu t \|\Vecs_{\Vecx}\|^2} H(t) e^{-s_t t} \mathrm{d}t {=} \int_{t \geq 0} \! e^{ - (s_t - \mu \|\Vecs_{\Vecx}\|^2) t} \mathrm{d}t \nonumber \\
											&= \frac{1}{s_t - \mu \|\Vecs_{\Vecx}\|^2},
\label{eq:LT_HeatKernel2}
\end{align}
provided $\Re{\left( s_t - {\mu}\|\Vecs_{\Vecx}\|^2 \right)} > 0 $, as required.
\subsection{Green's function of the three-dimensional wave equation}
\label{app:C6_WaveLT}
We now obtain the multidimensional bilateral Laplace transform of the Green's function for the wave equation \eqref{eq:WaveEqn}, which by definition must satisfy,
\begin{equation}
	\nabla^2 g(\Vecx, t) - \frac1{c^2} \frac{\partial^2}{\partial t^2}g(\Vecx, t)  = \delta(\Vecx, t).
\label{eq:GreenFunctionPDE}
\end{equation}
We begin by taking the Laplace transform of the PDE above, as follows:
\begin{align*}
	\displaystyle \int_{\Vecx \in \mathbb{R}^3}  \int_{t \in \mathbb{R}} \!\!\left( \nabla^2 g(\Vecx, t) {-} \frac1{c^2} \frac{\partial^2}{\partial t^2}g(\Vecx, t)  \right) \! e^{- (\Vecx,t)\cdot(\Vecs_{\Vecx}, s_t)} \mathrm{d}t \mathrm{d}\Vecx \\
	{=} \int_{\Vecx \in \mathbb{R}^3}  \int_{t \in \mathbb{R}} \delta(\Vecx, t)  e^{- (\Vecx,t)\cdot(\Vecs_{\Vecx}, s_t)}   \mathrm{d}t \mathrm{d}\Vecx 
\end{align*}
\begin{equation*}
\begin{split}
\displaystyle \int_{\Vecx \in \mathbb{R}^3} \left( \nabla^2 - \frac{s_t^2}{c^2} \right)  g_1(\Vecx, s_t)  e^{- \Vecx\cdot\Vecs_{\Vecx}} \mathrm{d}\Vecx
	= \int_{\Vecx \in \mathbb{R}^3}  \delta(\Vecx) \mathrm{d}\Vecx \\
	\Rightarrow \left( \|\Vecs_{\Vecx}\|^2 - ({s_t/c})^2 \right)G(\Vecs_{\Vecx},s_t)= 1.	
\end{split}
\end{equation*}
We can now rearrange to obtain the Laplace transform of $g$,
\begin{displaymath}
	G(\Vecs_{\Vecx},s_t) = \frac{1}{ \|\Vecs_{\Vecx}\|^2 - ({s_t/c})^2 }.
\end{displaymath}
%
%


\bibliographystyle{IEEEtran}
\bibliography{\MyBIBLocation/BIBfile_tsp2017}


\begin{IEEEbiography}[{\includegraphics[width=1in,height=1.25in,clip,keepaspectratio]{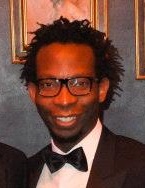}}]{John Murray-Bruce}
	is a Postdoctoral Research Associate with the Electrical and Computer Engineering Department at Boston University, Boston, MA since 2017. In 2012, he recieved the master's degree (first class) in electrical and electronic engineering from Imperial College, London and was awarded the Institute of Engineering and Technology (IET) Prize for 'best all round performance.' He received the Ph.D. degree in electrical and electronic engineering, in 2016, also from Imperial College, where he held a research assistant position in the Communications and Signal Processing Group from 2012 to 2017. His research interests are in interpolation, approximation and sampling theory, distributed algorithms, inverse problems of physical fields and computational imaging.
\end{IEEEbiography}
\begin{IEEEbiography}[{\includegraphics[width=1in,height=1.25in,clip,keepaspectratio]{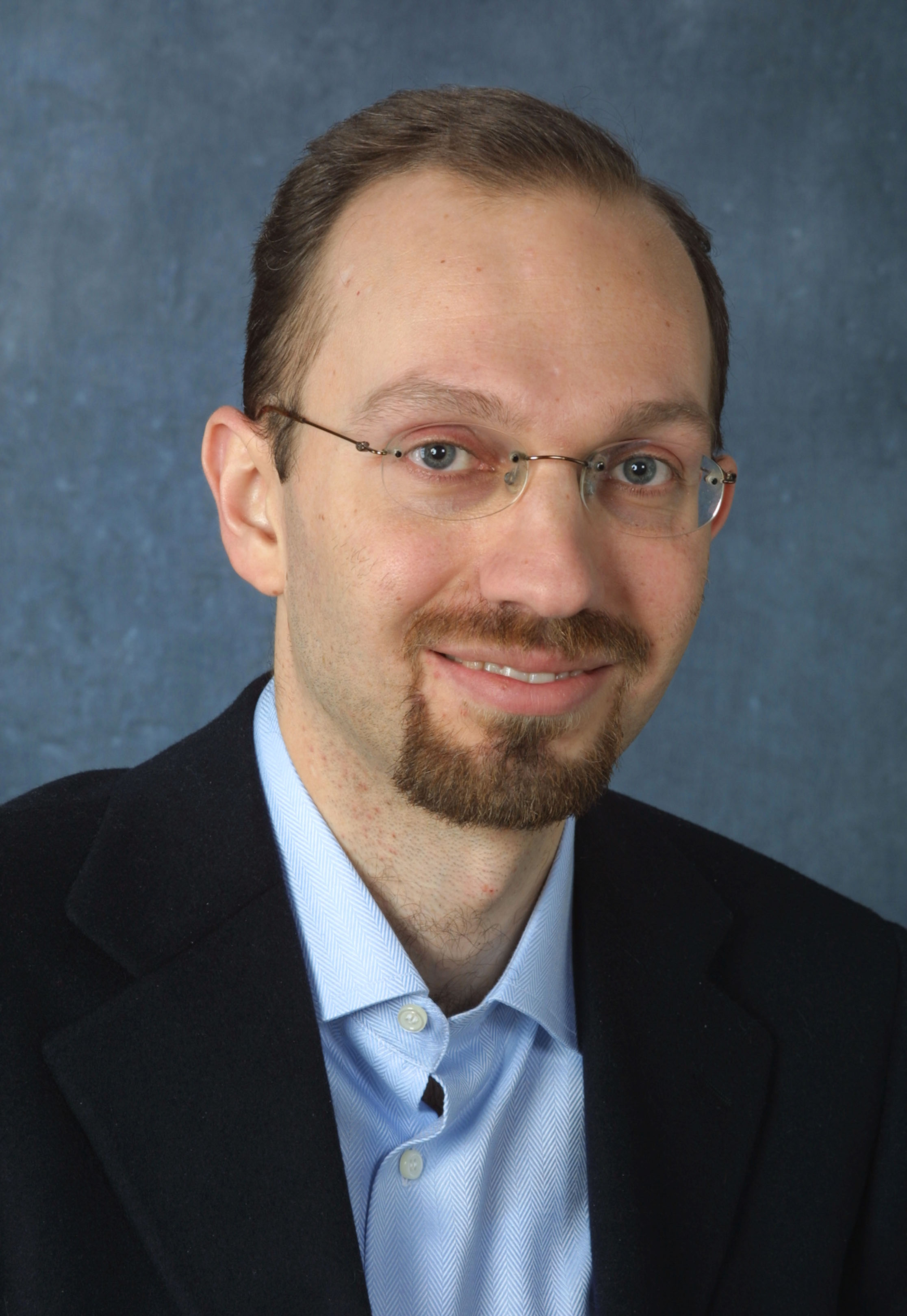}}]{Pier Luigi Dragotti}
	is Professor of Signal Processing with the Electrical and Electronic Engineering Department at Imperial College London. He received the Laurea degree (summa cum laude) in electrical and electronic engineering from the University of Naples Federico II, Naples, Italy, in 1997; the masters degree in communications systems from the Swiss Federal Institute of Technology of Lausanne (EPFL), Lausanne, Switzerland, in 1998, and the Ph.D. degree from EPFL in April 2002. He has held several visiting positions at different universities and research centers. He was a Visiting Student with Stanford University, Stanford, CA, USA, in 1996, a Researcher with the Mathematics of Communications Department, Bell Labs, Lucent Technologies, Murray Hill, NJ, USA, in 2000 and a Visiting Scientist with the Massachusetts Institute of Technology, Cambridge, MA, USA, in 2011. He was Technical Co-Chair for the European Signal Processing Conference in 2012, an Associate Editor of the IEEE TRANSACTIONS ON IMAGE PROCESSING from 2006 to 2009 and an Elected Member of the IEEE Image, Video and Multidimensional Signal Processing Technical Committee (2008-2013). Currently he is an elected member of the IEEE Signal Processing Theory and Method Technical Committee. He is a recipient of the ERC Starting Investigator Award for the project RecoSamp. His work includes sampling theory, wavelet theory and its applications, image and video compression, image-based rendering, and image super-resolution.
\end{IEEEbiography}

\balance
\end{document}